\documentclass[12pt,a4paper]{article}

\usepackage{color}
\usepackage{amscd}
\usepackage{amsmath}
\usepackage{psfrag}

\usepackage{authblk}
\usepackage{graphicx}
\usepackage{amsfonts}
\usepackage{pstricks,pst-node,pst-text,pst-3d}
\usepackage{tikz}
\usepackage{pgflibrarysnakes}
\usetikzlibrary[snakes]
\usetikzlibrary{arrows,shapes,positioning}
\usetikzlibrary{decorations.markings}
\tikzstyle arrowstyle=[scale=1]
\tikzstyle directed=[postaction={decorate,decoration={markings,
		mark=at position .65 with {\arrow[arrowstyle]{stealth}}}}]
\tikzstyle reverse directed=[postaction={decorate,decoration={markings,
		mark=at position .65 with {\arrowreversed[arrowstyle]{stealth};}}}]

\renewcommand{\baselinestretch}{1.5}

\newcommand{\im}{\mathrm i}

\newcommand{\tr}{\operatorname{Tr}}

\newcommand{\diag}{\operatorname{diag}}
\newcommand{\ket}[1]{\left|#1\right\rangle}      
\newcommand{\eq}{\begin{equation}}
	\newcommand{\en}{\end{equation}}
\newcommand{\bear}{\begin{eqnarray}}
	\newcommand{\ear}{\end{eqnarray}}

\def\weight[#1][#2][#3][#4][#5]{W
	\left(\left.\begin{array}{cc}
		#4&#3\\#1&#2\end{array}\right|\,#5
	\right)}

\def\weightT[#1][#2][#3][#4][#5]{\widetilde{W}
	\left(\left.\begin{array}{cc}
		#4&#3\\#1&#2\end{array}\right|\,#5
	\right)}

\title{\mbox{}Finite temperature properties of an integrable zigzag ladder chain}

\author{T.S. Tavares\footnote{E-mail: tavares@df.ufscar.br} \  \ and \ \ G.A.P. Ribeiro\footnote{E-mail: pavan@df.ufscar.br}}
\affil{\small Departamento de F\'{i}sica, Universidade Federal de S\~ao Carlos \\ S\~ao Carlos, SP 13565-905, Brazil}
\date{}

\begin{document}
\renewcommand{\baselinestretch}{1.2}
	
	\maketitle
	\thispagestyle{empty}
	
\begin{abstract}
We consider the interaction-round-a-face version of the six-vertex model for arbitrary anisotropy parameter, which allow us to derive an integrable one-dimensional quantum Hamiltonian with three-spin interactions. We apply the quantum transfer matrix approach for the face model version of the six-vertex model. The integrable quantum Hamiltonian shares some thermodynamical properties with the Heisenberg XXZ chain, but has different ordering and critical exponents. Two gapped phases are the dimerized antiferromagnetic order and the usual antiferromagnetic (N\'eel) order for positive nearest neighbour Ising coupling. In between these, there is an extended critical region, which is a quantum spin-liquid with broken parity symmetry inducing an oscillatory behavior at the long distance $<\sigma_1^z\sigma_{\ell+1}^z>$ correlation. At finite temperatures, the  numerical solution of the non-linear integral equations allows for the determination of the correlation length as well as for the momentum of the oscillation.
\end{abstract}

\newpage
\renewcommand{\baselinestretch}{1.5}

\section{Introduction}

The interaction-round-a-face (IRF) or just face models is a classical statistical mechanics model that was devised as a tool to solve the $8$-vertex model \cite{BAXTER1973,BAXTER,BAXTER1984}. Later on, several different realizations of face models were considered due to its own importance. In the realm of face models, there exist several well know models, like the solid-on-solid (SOS) type, the cyclic solid-on-solid (CSOS) \cite{BAXTER1973,BAXTER,BAXTER1984,KUNIBA,PEARCE1988} and the restricted solid-on-solid (RSOS) models \cite{PEARCE1989,PASQUIER,PEARCE1992,PEARCE1993a,PEARCE1993b}  and its $A$-$D$-$E$ generalizations \cite{PASQUIER,PEARCE1992,PEARCE1993a,PEARCE1993b}, which can be seen as a kind of face version of root of unity vertex models where the anisotropy parameter assumes only discrete values.

The connection between the face and vertex models has been largely exploited over the years. In fact this relation has been used to identify the operator content of the effective theory of the lattice model in the thermodynamic limit \cite{PEARCE1989}. On the vertex models side, there has been many instances of the study of quantum spin chain associated to the respective vertex model. The most notable example  is the relation between the six-vertex model and the Heisenberg (XXZ) quantum spin chain \cite{BAXTER,BOOK-KBI}. However, less was exploited from the direct relation of face models with its associated quantum spin chain counterpart.

In this work, we consider the face version of the six-vertex model for arbitrary anisotropy parameter, which goes along the same lines as done for the eight-vertex model \cite{BAXTER}. This allowed us to obtain
an integrable one dimensional quantum spin chain involving three-spin interaction terms. This spin chain can be arranged as a zigzag ladder. We extend the quantum transfer matrix approach from the context of vertex models to work in the face model setting. This allowed  the study of the physical properties of the spin chain in the thermodynamic limit. The IRF spin chain shares the same ground state diagram as the XXZ spin chain. Nevertheless, since the excited states over the ground state differ from the usual XXZ case, we obtained that the correlation length and operator content information are different.

This paper is organized as follows. In section \ref{integrability}, we described the general integrable structure of the face models and the respective formulation of the quantum transfer matrix approach in the face model context. In section \ref{IRFx6V}, we specialize the discussion to the face version of the six-vertex model, such that we can obtain the quantum spin chain and discuss its properties and the first few excited states. In section \ref{NLIE}, we introduce the non-linear integral equations that describes the ground state properties of the spin chain and discuss the solution of the equation at low temperature. In section \ref{res}, we discuss the phase diagram of the model and present some analytical and numerical solution of the non-linear integral equations for the correlation length. Our conclusions are given in section \ref{CONCLUSION}.

\section{The IRF model and the integrable Hamiltonian}\label{integrability}
The face models are classical statistical mechanical models on a square lattice defined by local Boltzmann weights, which can be depicted as \cite{BAXTER1973,BAXTER,BAXTER1984},
\eq
\begin{aligned}
\begin{tikzpicture}[scale=1.25]
\draw (0.0,0.) node {$\weight[a][b][c][d][\lambda]=$};
\draw (2.0,0) [-,color=black, thick]	+(-0.5,-0.5) -- +(-0.5,0.5)-- +(0.5,0.5) -- +(0.5,-0.5)-- +(-0.5,-0.5);
\draw (1.45,-0.66) node {$a$};
\draw (2.55,-0.66) node {$b$};
\draw (2.55, 0.64) node {$c$};
\draw (1.45, 0.65) node {$d$};
\draw (2., 0.) node {$\lambda$};
\draw (2.,0) [-,color=black,  thick, rounded corners=7pt]	+(-0.3,-0.5) -- +(-0.3,-0.3) -- +(-0.5,-0.3) ;
\draw (2.75, 0.) node {,};
\end{tikzpicture}
\end{aligned}
\en
where $a,b,c,d$ are the spins or heights of the corners of the face separated by bonds. The face weights depend on the spectral parameter $\lambda$.

The partition function of the classical $M\times L$ square lattice with periodic boundary condition
can be written as $Z_{\text{IRF}}=\tr{\left[\left(T_{\text{IRF}}(\lambda)\right)^M\right]}$, where $T_{\text{IRF}}(\lambda)$ is the transfer matrix, whose matrix elements are given by the product of the Boltzmann weights along the row such that,
\bear
T_{\text{IRF}}(\lambda)_{a_1 a_2 \cdots a_L}^{b_1 b_2 \cdots b_L}=\prod_{i=1}^L \weight[a_{i}][a_{i+1}][b_{i+1}][b_{i}][\lambda]\delta_{a_1 a_{L+1}}\delta_{b_1 b_{L+1}}.
\ear
The depiction of the transfer matrix is given in Figure \ref{transfermatrixFig}.
\begin{figure}[h]
	\begin{minipage}{\linewidth}
		\begin{center}
	\begin{tikzpicture}[scale=1.5]
		\draw (-0.650,0.) node {$T_{\text{IRF}}(\lambda)_{a_1 a_2 \cdots a_L}^{b_1 b_2 \cdots b_L}=$};
		\draw (1.0,0) [-,color=black, thick]	+(-0.5,-0.5) -- +(-0.5,0.5)-- +(0.5,0.5) -- +(0.5,-0.5)-- +(-0.5,-0.5);
		\draw (2.0,0) [-,color=black, thick]	+(-0.5,-0.5) -- +(-0.5,0.5)-- +(0.5,0.5) -- +(0.5,-0.5)-- +(-0.5,-0.5);
		\draw (3.0,0) [-,color=black, thick]	+(-0.5,-0.5) -- +(-0.5,0.5)-- +(0.5,0.5) -- +(0.5,-0.5)-- +(-0.5,-0.5);
		\draw (4.0,0) [-,color=black, thick]	+(-0.5,-0.5) -- +(-0.5,0.5)-- +(0.5,0.5) -- +(0.5,-0.5)-- +(-0.5,-0.5);
		\draw (5.0,0) [-,color=black, thick]	+(-0.5,-0.5) -- +(-0.5,0.5)-- +(0.5,0.5) -- +(0.5,-0.5)-- +(-0.5,-0.5);

		\draw (6.0,0) [-,color=black, thick]	+(-0.5,-0.5) -- +(-0.5,0.5)-- +(0.5,0.5) -- +(0.5,-0.5)-- +(-0.5,-0.5);
		
		\draw (1,0) [-,color=black,  rounded corners=7pt,thick]	+(-0.35,-0.5) -- +(-0.35,-0.35)-- +(-0.5,-0.35);
		\draw (2,0) [-,color=black,  rounded corners=7pt,thick]	+(-0.35,-0.5) -- +(-0.35,-0.35)-- +(-0.5,-0.35);
		\draw (3,0) [-,color=black,  rounded corners=7pt,thick]	+(-0.35,-0.5) -- +(-0.35,-0.35)-- +(-0.5,-0.35);
		\draw (4,0) [-,color=black,  rounded corners=7pt,thick]	+(-0.35,-0.5) -- +(-0.35,-0.35)-- +(-0.5,-0.35);
		\draw (6,0) [-,color=black,  rounded corners=7pt,thick]	+(-0.35,-0.5) -- +(-0.35,-0.35)-- +(-0.5,-0.35);
		
		\draw (0.45,-0.65) node {$a_1$};
		\draw (1.55,-0.65) node {$a_2$};
		\draw (2.55,-0.65) node {$a_3$};
		\draw (3.55,-0.65) node {$a_4$};
		\draw (4.55,-0.65) node {$a_5$};
		\draw (5.65,-0.65) node {$a_{L}$};
		\draw (6.75,-0.65) node {$a_{L+1}=a_{1}$};
		\draw (0.45, 0.75) node {$b_1$};
		\draw (1.55,0.75) node {$b_2$};
		\draw (2.55,0.75) node {$b_3$};
		\draw (3.55,0.75) node {$b_4$};
		\draw (4.55,0.75) node {$b_5$};
		\draw (5.65,0.75) node {$b_{L}$};
		\draw (6.75, 0.75) node {$b_{L+1}=b_{1}$};
		\draw (1., 0.) node {$\lambda$};
		\draw (2., 0.) node {$\lambda$};
		\draw (3., 0.) node {$\lambda$};
		\draw (4., 0.) node {$\lambda$};
		\draw (6., 0.) node {$\lambda$};
		\draw (5, 0.) node {$\cdots$};
	\end{tikzpicture}
\end{center}
\end{minipage}
\caption{Transfer matrix of the IRF model.}
\label{transfermatrixFig}
\end{figure}

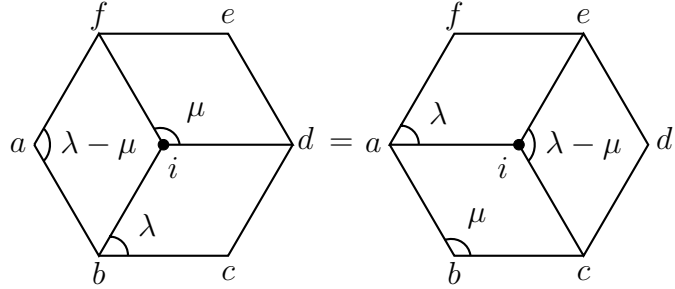
\begin{figure}[h]
\begin{minipage}{\linewidth}
\begin{center}
\begin{tikzpicture}[scale=0.85]

\draw (0,0) [-,color=black, thick]	+(0,0) -- +(1,1.73205)-- +(2,0) -- +(1,-1.73205)-- +(0,0);
\draw (0,0) [-,color=black, thick]	+(1,1.73205) -- +(3,1.73205)-- +(4,0) -- +(2,0);
\draw (0,0) [-,color=black, thick]	+(4,0) -- +(3,-1.73205)-- +(1,-1.73205);
\draw (-0.25,0) node {$a$};
\draw (1,-2) node {$b$};
\draw (3, -2) node {$c$};
\draw (4.2, 0.1) node {$d$};
\draw (3, 2) node {$e$};
\draw (1, 2) node {$f$};
\draw (2.15, -0.35) node {$i$};
\draw (2,0)[fill=black]  circle (0.45ex);

\draw (1., 0.) node {$\lambda-\mu$};
\draw (1.75,-1.25) node {$\lambda$};
\draw (2.5, 0.5) node {$\mu$};
\draw (0,0) [-,color=black,  thick, rounded corners=7pt]	+(0.15,-0.25) -- +(0.35,0.) -- +(0.15,0.25) ;
\draw (0,0) [-,color=black,  thick, rounded corners=7pt]	+(2.25,0) -- +(2.2,0.3) -- +(1.85,0.25) ;
\draw (0,0) [-,color=black,  thick, rounded corners=7pt]	+(1.45,-1.73205) -- +(1.45,-1.73205+0.3) -- +(1.2,-1.73205+0.3) ;

\draw (4.75, 0.) node {$=$};

\draw (7.5,0) [-,color=black, thick]	+(0,0) -- +(1,1.73205)-- +(2,0) -- +(1,-1.73205)-- +(0,0);
\draw (5.5,0) [-,color=black, thick]	+(2,0) -- +(0,0) -- +(1,1.73205) -- +(3,1.73205);
\draw (5.5,0) [-,color=black, thick]	+(3,-1.73205)-- +(1,-1.73205) -- +(0,0);

\draw (5.25,0) node {$a$};
\draw (6.5,-2) node {$b$};
\draw (8.5, -2) node {$c$};
\draw (9.75, 0.1) node {$d$};
\draw (8.5, 2) node {$e$};
\draw (6.5, 2) node {$f$};
\draw (7.25, -0.35) node {$i$};
\draw (7.5,0)[fill=black]  circle (0.45ex);

\draw (8.5, 0.) node {$\lambda-\mu$};
\draw (6.85,-1.15) node {$\mu$};
\draw (6.25, 0.5) node {$\lambda$};
\draw (7.5,0) [-,color=black,  thick, rounded corners=7pt]	+(0.15,-0.25) -- +(0.35,0.) -- +(0.15,0.25) ;
\draw (4.5,-1.73205) [-,color=black,  thick, rounded corners=7pt]	+(2.25,0) -- +(2.2,0.3) -- +(1.85,0.25) ;
\draw (4.5,1.73205) [-,color=black,  thick, rounded corners=7pt]	+(1.45,-1.73205) -- +(1.45,-1.73205+0.3) -- +(1.2,-1.73205+0.3) ;

\end{tikzpicture}
\end{center}
\end{minipage}
\caption{Yang-Baxter equation for IRF models. The black dot indicates that the index $i$ is summed over.}
\label{yangbaxter}
\end{figure}

The face model transfer matrix constitutes a family of commuting operators $[T_{\text{IRF}}(\lambda),T_{\text{IRF}}(\mu)]=0$ thanks to the Yang-Baxter equation. The face version of the
Yang-Baxter reads,
\bear
\sum_{i} \weight[a][b][i][f][\lambda-\mu]\weight[i][d][e][f][\mu]\weight[b][c][d][i][\lambda]\nonumber\\
=\sum_{i} \weight[a][i][e][f][\lambda]\weight[b][c][i][a][\mu]\weight[i][c][d][e][\lambda-\mu],
\ear
which again can be displayed in graphical notation as in Figure \ref{yangbaxter}.

\begin{figure}[h]
	\begin{minipage}{\linewidth}
		\begin{center}
			\begin{tikzpicture}[scale=1.1]
				
				\draw (0,0) [-,color=black, thick]	+(0,0) -- +(1,1)-- +(2,0) -- +(1,-1)-- +(0,0);
				\draw (2,0) [-,color=black, thick]	+(0,0) -- +(1,1)-- +(2,0) -- +(1,-1)-- +(0,0);
				\draw (0,0) [-,color=black, dotted]	+(1,-1) -- +(3,-1);
				\draw (0,0) [-,color=black, dotted]	+(1,1) -- +(3,1);
				\draw (-0.2,0) node {$a$};
				\draw (1,-1.25) node {$b$};
				\draw (3, -1.25) node {$b$};
				\draw (4.2, 0) node {$c$};
				\draw (3, 1.25) node {$d$};
				\draw (1, 1.25) node {$d$};
				\draw (2., -0.35) node {$i$};
				\draw (2,0)[fill=black]  circle (0.45ex);
				
				\draw (0.5, 0.) node {$\lambda$};
				\draw (2.55,0) node {$-\lambda$};
				\draw (0,0) [-,color=black,  thick, rounded corners=7pt]	+(0.15,-0.15) -- +(0.3,0.) -- +(0.15,0.15) ;
				\draw (2,0) [-,color=black,  thick, rounded corners=7pt]	+(0.15,-0.15) -- +(0.3,0.) -- +(0.15,0.15);

				\draw (4.55, 0.) node {$=$};

				\draw (5.3,0) [-,color=black, thick]	+(0,0) -- +(1,1);
				\draw (5.3,0) [-,color=black, thick]	+(0,0) -- +(1,-1);
				\draw (8.3,0) [-,color=black, thick]	+(0,1) -- +(1,0);
				\draw (8.3,0) [-,color=black, thick]	+(0,-1) -- +(1,0);
				\draw (5.3,0) [-,color=black, dotted]	+(0,0) -- +(4,0);
				\draw (5.3,0) [-,color=black, dotted]	+(1,-1) -- +(3,-1);
				\draw (5.3,0) [-,color=black, dotted]	+(1,1) -- +(3,1);

				\draw (5.05,0) node {$a$};
				\draw (6.3,-1.25) node {$b$};
				\draw (8.3, -1.25) node {$b$};
				\draw (6.3, 1.25) node {$d$};
				\draw (8.3, 1.25) node {$d$};
				\draw (9.45, 0) node {$c$};
				\draw (11.0, 0) node {$\rho(\lambda)\rho(-\lambda)\delta_{ac}$};

			\end{tikzpicture}
		\end{center}
	\end{minipage}
	\caption{Unitarity condition.}
	\label{unitarity}
\end{figure}
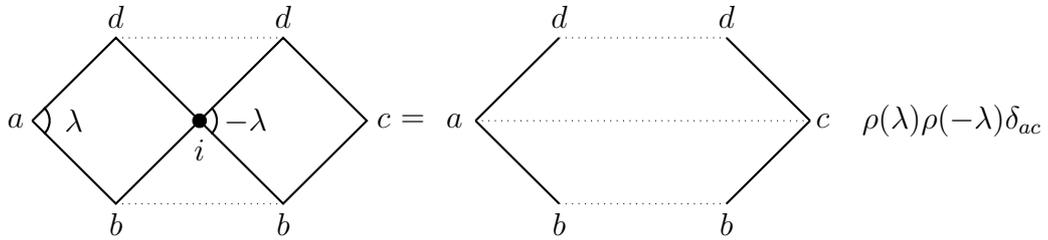

Besides the Yang-Baxter equation, the face weights satisfy a number of important properties, such as the  unitarity,
\bear
\sum_i \weight[a][b][i][d][\lambda]\weight[i][b][c][d][-\lambda]=\rho(\lambda)\rho(-\lambda)\delta_{ac},
\ear
which is graphically depicted as in Figure \ref{unitarity}.

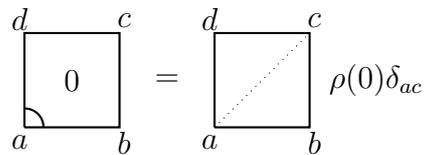
\begin{figure}[h]
\begin{center}
	\begin{tikzpicture}[scale=1.25]
		\draw (2.0,0) [-,color=black, thick]	+(-0.5,-0.5) -- +(-0.5,0.5)-- +(0.5,0.5) -- +(0.5,-0.5)-- +(-0.5,-0.5);
		\draw (1.45,-0.66) node {$a$};
		\draw (2.55,-0.66) node {$b$};
		\draw (2.55, 0.64) node {$c$};
		\draw (1.45, 0.65) node {$d$};
		\draw (2., 0.) node {$0$};
		\draw (3., 0.) node {$=$};
		\draw (2.,0) [-,color=black,  thick, rounded corners=7pt]	+(-0.3,-0.5) -- +(-0.3,-0.3) -- +(-0.5,-0.3) ;

		\draw (4.0,0) [-,color=black, thick]	+(-0.5,-0.5) -- +(-0.5,0.5)-- +(0.5,0.5) -- +(0.5,-0.5)-- +(-0.5,-0.5);
		\draw (4.0,0) [-,color=black, dotted]	+(-0.5,-0.5) -- +(0.5,0.5);
		\draw (3.45,-0.66) node {$a$};
		\draw (4.55,-0.66) node {$b$};
		\draw (4.55, 0.64) node {$c$};
		\draw (3.45, 0.65) node {$d$};
		\draw (5.2, 0) node {$\rho(0)\delta_{ac}$};
	\end{tikzpicture}
\end{center}
\caption{Initial condition.}
\label{initial}
\end{figure}

In addition, one has the initial condition given as,
\eq
\weight[a][b][c][d][0]=\rho(0)\delta_{ac}.
\en
which is depicted in Figure \ref{initial}.
The face model transfer matrix $T_{\text{IRF}}(\lambda)$  can also be seen as a generating function of conserved charges. For instance, by use of the above properties of the face weights, one can obtain a quantum one-dimensional Hamiltonian via the logarithmic derivative of the transfer matrix ${\cal H}_{\text{IRF}}=\partial_{\lambda}\log T_{\text{IRF}}(\lambda) |_{\lambda=0}$, whose face representation is given below (See Figure \ref{hamiltonian}).
\begin{figure}[h]
\begin{center}
	\begin{tikzpicture}[scale=1.5]
		\draw (-0.55,0.) node {$\displaystyle {\cal H}_{\text{IRF}}=\sum_{i=1}^L \partial_{\lambda=0}$};
		\draw (1.0,0) [-,color=black, thick]	+(-0.5,0) -- +(0,0.5)-- +(0.5,0) -- +(0,-0.5)-- +(-0.5,0);
		\draw (1.0,0) [-,color=black, dotted]	+(-0.5,-0.75) -- +(-0.5,0.75);
		\draw (1.0,0) [-,color=black, dotted]	+(0,0.5) -- +(0,0.75);
		\draw (2.0,0) [-,color=black, dotted]	+(-0.5,-0.75) -- +(-0.5,0.75);
		\draw (1.0,0) [-,color=black, dotted]	+(0,-0.5) -- +(0,-0.75);
		\draw (1,-1) node {$i$};
		\draw (0.5,-1) node {$i-1$};
		\draw (1.5,-1) node {$i+1$};
		\draw (1., 0.) node {$\lambda$};
		\draw (1.,0) [-,color=black,  thick, rounded corners=7pt]	+(0.15,-0.35) -- +(0,-0.25) -- +(-0.15,-0.35) ;
	\end{tikzpicture}
\end{center}
\caption{Face representation of the one-dimensional quantum Hamiltonian.}
\label{hamiltonian}
\end{figure}

Proceeding along the same lines as in the vertex model case  \cite{KLUMPER92}, one can define the partition function of the one-dimensional quantum spin chain as $Z_L=\tr\left[e^{-\beta {\cal H}_{\text{IRF}}} \right]$, which can be mapped via the Trotter-Suzuki decomposition into a special $N\times L$ classical lattice, where $N$ is the so-called Trotter number and $\beta=1/T$ is the inverse of temperature. This is done by introducing a conjugated face transfer matrix, $\overline{T}_{\text{IRF}}(\lambda)$,
\bear
\overline{T}_{\text{IRF}}(\lambda)_{a_1 a_2 \cdots a_L}^{b_1 b_2 \cdots b_L}=\prod_{i=1}^L \weightT[a_{i}][a_{i+1}][b_{i+1}][b_{i}][\lambda]\delta_{a_1 a_{L+1}}\delta_{b_1 b_{L+1}},
\ear
which is simply depicted as in Figure \ref{conjugatedT}.
\begin{figure}[h]
	\begin{center}
		\begin{tikzpicture}[scale=1.5]
	\draw (-0.650,0.) node {$\overline{T}_{\text{IRF}}(\lambda)_{a_1 a_2 \cdots a_L}^{b_1 b_2 \cdots b_L}=$};
	\draw (1.0,0) [-,color=black, thick]	+(-0.5,-0.5) -- +(-0.5,0.5)-- +(0.5,0.5) -- +(0.5,-0.5)-- +(-0.5,-0.5);
	\draw (2.0,0) [-,color=black, thick]	+(-0.5,-0.5) -- +(-0.5,0.5)-- +(0.5,0.5) -- +(0.5,-0.5)-- +(-0.5,-0.5);
	\draw (3.0,0) [-,color=black, thick]	+(-0.5,-0.5) -- +(-0.5,0.5)-- +(0.5,0.5) -- +(0.5,-0.5)-- +(-0.5,-0.5);
	\draw (4.0,0) [-,color=black, thick]	+(-0.5,-0.5) -- +(-0.5,0.5)-- +(0.5,0.5) -- +(0.5,-0.5)-- +(-0.5,-0.5);
	\draw (5.0,0) [-,color=black, thick]	+(-0.5,-0.5) -- +(-0.5,0.5)-- +(0.5,0.5) -- +(0.5,-0.5)-- +(-0.5,-0.5);

	\draw (6.0,0) [-,color=black, thick]	+(-0.5,-0.5) -- +(-0.5,0.5)-- +(0.5,0.5) -- +(0.5,-0.5)-- +(-0.5,-0.5);
	
	\draw (1,0) [-,color=black,  rounded corners=7pt,thick]	+(0.35,-0.5) -- +(0.35,-0.35)-- +(0.5,-0.35);
	\draw (2,0) [-,color=black,  rounded corners=7pt,thick]	+(0.35,-0.5) -- +(0.35,-0.35)-- +(0.5,-0.35);
	\draw (3,0) [-,color=black,  rounded corners=7pt,thick]	+(0.35,-0.5) -- +(0.35,-0.35)-- +(0.5,-0.35);
	\draw (4,0) [-,color=black,  rounded corners=7pt,thick]	+(0.35,-0.5) -- +(0.35,-0.35)-- +(0.5,-0.35);
	\draw (6,0) [-,color=black,  rounded corners=7pt,thick]	+(0.35,-0.5) -- +(0.35,-0.35)-- +(0.5,-0.35);

	\draw (0.45,-0.65) node {$a_1$};
	\draw (1.55,-0.65) node {$a_2$};
	\draw (2.55,-0.65) node {$a_3$};
	\draw (3.55,-0.65) node {$a_4$};
	\draw (4.55,-0.65) node {$a_5$};
	\draw (5.65,-0.65) node {$a_{L}$};
	\draw (6.75,-0.65) node {$a_{L+1}=a_{1}$};
	\draw (0.45, 0.75) node {$b_1$};
	\draw (1.55,0.75) node {$b_2$};
	\draw (2.55,0.75) node {$b_3$};
	\draw (3.55,0.75) node {$b_4$};
	\draw (4.55,0.75) node {$b_5$};
	\draw (5.65,0.75) node {$b_{L}$};
	\draw (6.75, 0.75) node {$b_{L+1}=b_{1}$};
	\draw (1., 0.) node {$\lambda$};
	\draw (2., 0.) node {$\lambda$};
	\draw (3., 0.) node {$\lambda$};
	\draw (4., 0.) node {$\lambda$};
	\draw (6., 0.) node {$\lambda$};
	\draw (5, 0.) node {$\cdots$};
		\end{tikzpicture}
	\end{center}
	\caption{Conjugated transfer matrix such that $\overline{T}_{\text{IRF}}(\lambda) \sim \overline{T}_{\text{IRF}}(0) e^{\lambda {\cal H}_{\text{IRF}}+ O(\lambda^2)}$.}
	\label{conjugatedT}
\end{figure}

\begin{figure}[h]
	\begin{center}
		\begin{tikzpicture}[scale=1.5]
			\draw (-0.650,0.) node {$T_{\text{IRF}}(-\tfrac{\beta }{N})$};
			\draw (1.0,0) [-,color=black, thick]	+(-0.5,-0.5) -- +(-0.5,0.5)-- +(0.5,0.5) -- +(0.5,-0.5)-- +(-0.5,-0.5);
			\draw (2.0,0) [-,color=black, thick]	+(-0.5,-0.5) -- +(-0.5,0.5)-- +(0.5,0.5) -- +(0.5,-0.5)-- +(-0.5,-0.5);
			\draw (3.0,0) [-,color=black, thick]	+(-0.5,-0.5) -- +(-0.5,0.5)-- +(0.5,0.5) -- +(0.5,-0.5)-- +(-0.5,-0.5);
			\draw (4.0,0) [-,color=black, thick]	+(-0.5,-0.5) -- +(-0.5,0.5)-- +(0.5,0.5) -- +(0.5,-0.5)-- +(-0.5,-0.5);
			\draw (5.0,0) [-,color=black, thick]	+(-0.5,-0.5) -- +(-0.5,0.5)-- +(0.5,0.5) -- +(0.5,-0.5)-- +(-0.5,-0.5);
			
			\draw (6.0,0) [-,color=black, thick]	+(-0.5,-0.5) -- +(-0.5,0.5)-- +(0.5,0.5) -- +(0.5,-0.5)-- +(-0.5,-0.5);
			
			\draw (1,0) [-,color=black,  rounded corners=7pt,thick]	+(-0.35,-0.5) -- +(-0.35,-0.35)-- +(-0.5,-0.35);
			\draw (2,0) [-,color=black,  rounded corners=7pt,thick]	+(-0.35,-0.5) -- +(-0.35,-0.35)-- +(-0.5,-0.35);
			\draw (3,0) [-,color=black,  rounded corners=7pt,thick]	+(-0.35,-0.5) -- +(-0.35,-0.35)-- +(-0.5,-0.35);
			\draw (4,0) [-,color=black,  rounded corners=7pt,thick]	+(-0.35,-0.5) -- +(-0.35,-0.35)-- +(-0.5,-0.35);
			\draw (6,0) [-,color=black,  rounded corners=7pt,thick]	+(-0.35,-0.5) -- +(-0.35,-0.35)-- +(-0.5,-0.35);

			\draw (-0.650,1.) node {$\overline{T}_{\text{IRF}}(-\tfrac{\beta }{N})$};
			\draw (1.0,1) [-,color=black, thick]	+(-0.5,-0.5) -- +(-0.5,0.5)-- +(0.5,0.5) -- +(0.5,-0.5)-- +(-0.5,-0.5);
			\draw (2.0,1) [-,color=black, thick]	+(-0.5,-0.5) -- +(-0.5,0.5)-- +(0.5,0.5) -- +(0.5,-0.5)-- +(-0.5,-0.5);
			\draw (3.0,1) [-,color=black, thick]	+(-0.5,-0.5) -- +(-0.5,0.5)-- +(0.5,0.5) -- +(0.5,-0.5)-- +(-0.5,-0.5);
			\draw (4.0,1) [-,color=black, thick]	+(-0.5,-0.5) -- +(-0.5,0.5)-- +(0.5,0.5) -- +(0.5,-0.5)-- +(-0.5,-0.5);
			\draw (5.0,1) [-,color=black, thick]	+(-0.5,-0.5) -- +(-0.5,0.5)-- +(0.5,0.5) -- +(0.5,-0.5)-- +(-0.5,-0.5);
			\draw (6.0,1) [-,color=black, thick]	+(-0.5,-0.5) -- +(-0.5,0.5)-- +(0.5,0.5) -- +(0.5,-0.5)-- +(-0.5,-0.5);
			
			\draw (1,1) [-,color=black,  rounded corners=7pt,thick]	+(0.35,-0.5) -- +(0.35,-0.35)-- +(0.5,-0.35);
			\draw (2,1) [-,color=black,  rounded corners=7pt,thick]	+(0.35,-0.5) -- +(0.35,-0.35)-- +(0.5,-0.35);
			\draw (3,1) [-,color=black,  rounded corners=7pt,thick]	+(0.35,-0.5) -- +(0.35,-0.35)-- +(0.5,-0.35);
			\draw (4,1) [-,color=black,  rounded corners=7pt,thick]	+(0.35,-0.5) -- +(0.35,-0.35)-- +(0.5,-0.35);
			\draw (6,1) [-,color=black,  rounded corners=7pt,thick]	+(0.35,-0.5) -- +(0.35,-0.35)-- +(0.5,-0.35);

			\draw (-0.650,2.) node {$\vdots$};
			\draw (1.0,2) [-,color=black, thick]	+(-0.5,-0.5) -- +(-0.5,0.5)-- +(0.5,0.5) -- +(0.5,-0.5)-- +(-0.5,-0.5);
			\draw (2.0,2) [-,color=black, thick]	+(-0.5,-0.5) -- +(-0.5,0.5)-- +(0.5,0.5) -- +(0.5,-0.5)-- +(-0.5,-0.5);
			\draw (3.0,2) [-,color=black, thick]	+(-0.5,-0.5) -- +(-0.5,0.5)-- +(0.5,0.5) -- +(0.5,-0.5)-- +(-0.5,-0.5);
			\draw (4.0,2) [-,color=black, thick]	+(-0.5,-0.5) -- +(-0.5,0.5)-- +(0.5,0.5) -- +(0.5,-0.5)-- +(-0.5,-0.5);
			\draw (5.0,2) [-,color=black, thick]	+(-0.5,-0.5) -- +(-0.5,0.5)-- +(0.5,0.5) -- +(0.5,-0.5)-- +(-0.5,-0.5);
			
			\draw (6.0,2) [-,color=black, thick]	+(-0.5,-0.5) -- +(-0.5,0.5)-- +(0.5,0.5) -- +(0.5,-0.5)-- +(-0.5,-0.5);
			
			\draw (-0.650,3.) node {$T_{\text{IRF}}(-\tfrac{\beta }{N})$};
			\draw (1,3) [-,color=black,  rounded corners=7pt,thick]	+(-0.35,-0.5) -- +(-0.35,-0.35)-- +(-0.5,-0.35);
			\draw (2,3) [-,color=black,  rounded corners=7pt,thick]	+(-0.35,-0.5) -- +(-0.35,-0.35)-- +(-0.5,-0.35);
			\draw (3,3) [-,color=black,  rounded corners=7pt,thick]	+(-0.35,-0.5) -- +(-0.35,-0.35)-- +(-0.5,-0.35);
			\draw (4,3) [-,color=black,  rounded corners=7pt,thick]	+(-0.35,-0.5) -- +(-0.35,-0.35)-- +(-0.5,-0.35);
			\draw (6,3) [-,color=black,  rounded corners=7pt,thick]	+(-0.35,-0.5) -- +(-0.35,-0.35)-- +(-0.5,-0.35);
			
			\draw (-0.650,4.) node {$\overline{T}_{\text{IRF}}(-\tfrac{\beta }{N})$};
			\draw (1.0,3) [-,color=black, thick]	+(-0.5,-0.5) -- +(-0.5,0.5)-- +(0.5,0.5) -- +(0.5,-0.5)-- +(-0.5,-0.5);
			\draw (2.0,3) [-,color=black, thick]	+(-0.5,-0.5) -- +(-0.5,0.5)-- +(0.5,0.5) -- +(0.5,-0.5)-- +(-0.5,-0.5);
			\draw (3.0,3) [-,color=black, thick]	+(-0.5,-0.5) -- +(-0.5,0.5)-- +(0.5,0.5) -- +(0.5,-0.5)-- +(-0.5,-0.5);
			\draw (4.0,3) [-,color=black, thick]	+(-0.5,-0.5) -- +(-0.5,0.5)-- +(0.5,0.5) -- +(0.5,-0.5)-- +(-0.5,-0.5);
			\draw (5.0,3) [-,color=black, thick]	+(-0.5,-0.5) -- +(-0.5,0.5)-- +(0.5,0.5) -- +(0.5,-0.5)-- +(-0.5,-0.5);
			\draw (6.0,3) [-,color=black, thick]	+(-0.5,-0.5) -- +(-0.5,0.5)-- +(0.5,0.5) -- +(0.5,-0.5)-- +(-0.5,-0.5);
			
			\draw (1,4) [-,color=black,  rounded corners=7pt,thick]	+(0.35,-0.5) -- +(0.35,-0.35)-- +(0.5,-0.35);
			\draw (2,4) [-,color=black,  rounded corners=7pt,thick]	+(0.35,-0.5) -- +(0.35,-0.35)-- +(0.5,-0.35);
			\draw (3,4) [-,color=black,  rounded corners=7pt,thick]	+(0.35,-0.5) -- +(0.35,-0.35)-- +(0.5,-0.35);
			\draw (4,4) [-,color=black,  rounded corners=7pt,thick]	+(0.35,-0.5) -- +(0.35,-0.35)-- +(0.5,-0.35);
			\draw (6,4) [-,color=black,  rounded corners=7pt,thick]	+(0.35,-0.5) -- +(0.35,-0.35)-- +(0.5,-0.35);
			
			\draw (1.0,4) [-,color=black, thick]	+(-0.5,-0.5) -- +(-0.5,0.5)-- +(0.5,0.5) -- +(0.5,-0.5)-- +(-0.5,-0.5);
			\draw (2.0,4) [-,color=black, thick]	+(-0.5,-0.5) -- +(-0.5,0.5)-- +(0.5,0.5) -- +(0.5,-0.5)-- +(-0.5,-0.5);
			\draw (3.0,4) [-,color=black, thick]	+(-0.5,-0.5) -- +(-0.5,0.5)-- +(0.5,0.5) -- +(0.5,-0.5)-- +(-0.5,-0.5);
			\draw (4.0,4) [-,color=black, thick]	+(-0.5,-0.5) -- +(-0.5,0.5)-- +(0.5,0.5) -- +(0.5,-0.5)-- +(-0.5,-0.5);
			\draw (5.0,4) [-,color=black, thick]	+(-0.5,-0.5) -- +(-0.5,0.5)-- +(0.5,0.5) -- +(0.5,-0.5)-- +(-0.5,-0.5);
			\draw (6.0,4) [-,color=black, thick]	+(-0.5,-0.5) -- +(-0.5,0.5)-- +(0.5,0.5) -- +(0.5,-0.5)-- +(-0.5,-0.5);

			\draw (5, 0.) node {$\cdots$};
			\draw (5, 1.) node {$\cdots$};
			\draw (5, 3.) node {$\cdots$};
			\draw (5, 4.) node {$\cdots$};
			\draw (1., 2.) node {$\vdots$};
			\draw (2., 2.) node {$\vdots$};
			\draw (3., 2.) node {$\vdots$};
			\draw (4., 2.) node {$\vdots$};
			\draw (6., 2.) node {$\vdots$};

			\draw (1, -0.75) node {$1$};
			\draw (2, -0.75) node {$2$};
			\draw (6, -0.75) node {$L$};
			
			\draw (0.35, 0) node {$1$};
			\draw (0.35, 1) node {$2$};
			\draw (0.35, 4) node {$N$};

	\end{tikzpicture}
	\end{center}
	\caption{Trotter-Suzuki decomposition for face models.}
	\label{trotter}
\end{figure}
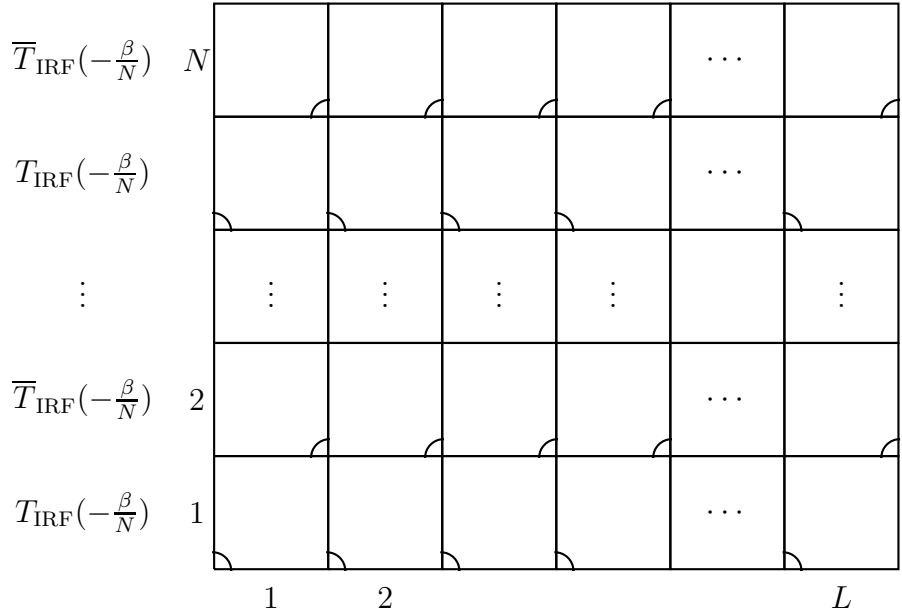

In terms of both face transfer matrices, the partition function can be written as
\eq
Z_L=\tr\left[e^{-\beta{\cal H}_{\text{IRF}}}\right]=\lim_{N\to\infty}\tr\left[ \left(T_{\text{IRF}}(-\tfrac{\beta }{N})\overline{T}_{\text{IRF}}(-\tfrac{\beta }{N})\right)^{N/2} \right],
\en
as shown in Figure \ref{trotter}.

In this context, it is convenient to introduce another transfer matrix along the vertical direction. This transfer matrix is usually called the quantum transfer matrix (QTM) \cite{KLUMPER92}, which can be defined as follows,
\begin{align}
&T^{\text{QTM}}(x)_{a_1 a_2 \cdots a_N}^{b_1 b_2 \cdots b_N}=\\
&=\prod_{i=1}^{N/2} \weight[a_{2i-1}][a_{2i}][b_{2i}][b_{2i-1}][\im x+\lambda]\weightT[a_{2i}][a_{2i+1}][b_{2i+1}][b_{2i}][-\im x+\lambda]\delta_{a_1 a_{N+1}}\delta_{b_1 b_{N+1}}. \nonumber
\end{align}
where $\lambda=-\frac{\beta}{ N}$.

In general, the investigation of the quantum transfer matrix eigenvalues determines the physical properties of the one-dimensional Hamiltonian. By means of the quantum transfer matrix approach one can obtain a set of non-linear integral equations\cite{KLUMPER92}, which yields the thermodynamical properties of the model.

\section{The face version of the six and eight-vertex model}\label{IRFx6V}

In this section, we consider the face version of the six-vertex model for arbitrary anisotropy parameter, along the same lines as done for the eight-vertex model in \cite{BAXTER}. This allows us to obtain an integrable one-dimensional quantum spin chain and by means of the quantum transfer matrix approach we can study its physical properties in the thermodynamic limit at finite temperatures via a set of non-linear integral equations. Similarly, we could set up non-linear integral equations for finite chain length $L$ at zero temperature, this would change only the driving term of the equations \cite{KLUMPER93}.

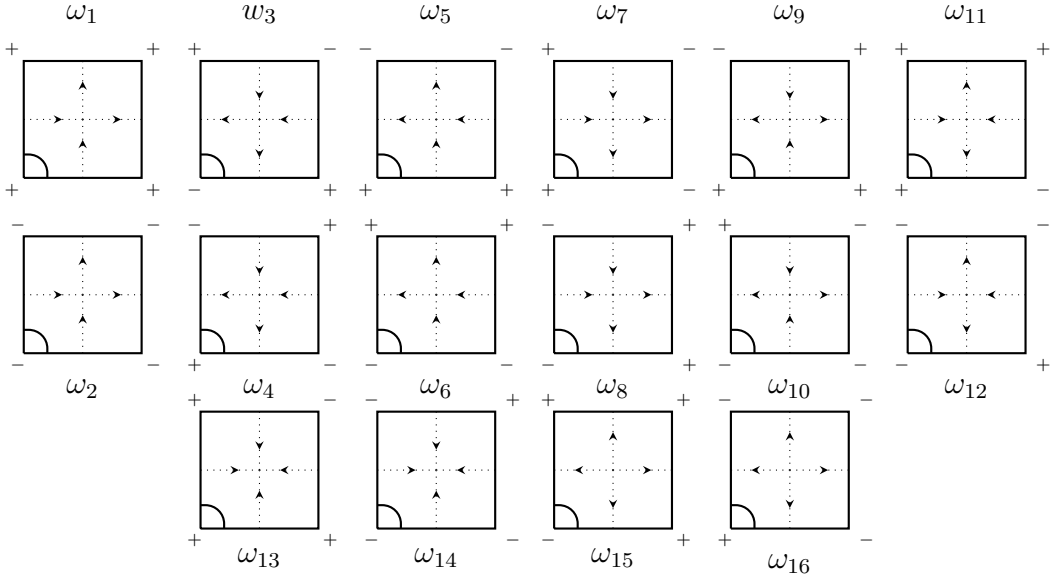
\begin{figure}[tb]
\begin{minipage}{\linewidth}
\begin{center}
\begin{tikzpicture}[scale=1.55]
\draw (0,1.5) [-,color=black, dotted, directed, rounded corners=7pt]+(-(0.,0) --+(0.,0.5);
\draw (0,1.5) [-,color=black, dotted, directed, rounded corners=7pt]+(-(0.,-0.5) --+(0.,0.);
\draw (0,1.5) [-,color=black,  dotted, directed, rounded corners=7pt]	+(0,0) -- +(0.5,0) ;
\draw (0,1.5) [-,color=black,  dotted, directed, rounded corners=7pt]	+(-0.5,0) -- +(0.,0) ;
\draw (0,1.5) [-,color=black,  thick, rounded corners=7pt]	+(-0.3,-0.5) -- +(-0.3,-0.3) -- +(-0.5,-0.3) ;
\draw (0,1.5) [-,color=black, thick]	+(-0.5,-0.5) -- +(-0.5,0.5)-- +(0.5,0.5) -- +(0.5,-0.5)--+(-0.5,-0.5);
\draw (-0.6,-0.6+1.5) node {\tiny $+$};
\draw (0.6,-0.6+1.5) node {\tiny $+$};
\draw (0.6, 0.6+1.5) node {\tiny$+$};
\draw (-0.6, 0.6+1.5) node {\tiny$+$};
\draw (0, 0.9+1.5) node {$\omega_1$};

\draw (1.5,1.5) [-,color=black, dotted, directed, rounded corners=7pt]+(-(0.,0.5) --+(0.,0);
\draw (1.5,1.5) [-,color=black, dotted, directed, rounded corners=7pt]+(-(0.,0) --+(0.,-0.5);
\draw (1.5,1.5) [-,color=black,  dotted, directed, rounded corners=7pt]	+(0.5,0) -- +(0,0) ;
\draw (1.5,1.5) [-,color=black,  dotted, directed, rounded corners=7pt]	+(0,0) -- +(-0.5,0) ;
\draw (1.5,1.5) [-,color=black,  thick, rounded corners=7pt]	+(-0.3,-0.5) -- +(-0.3,-0.3) -- +(-0.5,-0.3) ;
\draw (1.5,1.5) [-,color=black, thick]	+(-0.5,-0.5) -- +(-0.5,0.5)-- +(0.5,0.5) -- +(0.5,-0.5)--+(-0.5,-0.5);
\draw (0.95,-0.6+1.5) node {\tiny$-$};
\draw (2.1,-0.6+1.5) node {\tiny$+$};
\draw (2.1, 0.6+1.5) node {\tiny$-$};
\draw (0.95, 0.6+1.5) node {\tiny$+$};
\draw (1.5, 0.9+1.5) node {$w_3$};

\draw (0,0) [-,color=black, dotted, directed, rounded corners=7pt]+(-(0.,0) --+(0.,0.5);
\draw (0,0) [-,color=black, dotted, directed, rounded corners=7pt]+(-(0.,-0.5) --+(0.,0);
\draw (0,0) [-,color=black,  dotted, directed, rounded corners=7pt]	+(0,0) -- +(0.5,0) ;
\draw (0,0) [-,color=black,  dotted, directed, rounded corners=7pt]	+(-0.5,0) -- +(0,0) ;
\draw (0,0) [-,color=black,  thick, rounded corners=7pt]	+(-0.3,-0.5) -- +(-0.3,-0.3) -- +(-0.5,-0.3) ;
\draw (0,0) [-,color=black, thick]	+(-0.5,-0.5) -- +(-0.5,0.5)-- +(0.5,0.5) -- +(0.5,-0.5)--+(-0.5,-0.5);
\draw (-0.55,-0.6) node {\tiny $-$};
\draw (0.6,-0.6) node {\tiny $-$};
\draw (0.6, 0.6) node {\tiny $-$};
\draw (-0.55, 0.6) node {\tiny $-$};
\draw (0, -0.8) node {$\omega_2$};

\draw (1.5,0) [-,color=black, dotted, directed, rounded corners=7pt]+(-(0.,0.5) --+(0.,0);
\draw (1.5,0) [-,color=black, dotted, directed, rounded corners=7pt]+(-(0.,0) --+(0.,-0.5);
\draw (1.5,0) [-,color=black,  dotted, directed, rounded corners=7pt]	+(0.5,0) -- +(0,0) ;
\draw (1.5,0) [-,color=black,  dotted, directed, rounded corners=7pt]	+(0,0) -- +(-0.5,0) ;
\draw (1.5,0) [-,color=black,  thick, rounded corners=7pt]	+(-0.3,-0.5) -- +(-0.3,-0.3) -- +(-0.5,-0.3) ;
\draw (1.5,0) [-,color=black, thick]	+(-0.5,-0.5) -- +(-0.5,0.5)-- +(0.5,0.5) -- +(0.5,-0.5)--+(-0.5,-0.5);
\draw (0.95,-0.6) node {\tiny$+$};
\draw (2.1,-0.6) node {\tiny$-$};
\draw (2.1, 0.6) node {\tiny$+$};
\draw (0.95, 0.6) node {\tiny$-$};
\draw (1.5, -0.8) node {$\omega_4$};


\draw (3,1.5) [-,color=black, dotted, directed, rounded corners=7pt]+(-(0.,0) --+(0.,0.5);
\draw (3,1.5) [-,color=black, dotted, directed, rounded corners=7pt]+(-(0.,-0.5) --+(0.,0.);
\draw (3,1.5) [-,color=black,  dotted, directed, rounded corners=7pt]	+(0.5,0) -- +(0,0) ;
\draw (3,1.5) [-,color=black,  dotted, directed, rounded corners=7pt]	+(0,0) -- +(-0.5,0) ;
\draw (3,1.5) [-,color=black,  thick, rounded corners=7pt]	+(-0.3,-0.5) -- +(-0.3,-0.3) -- +(-0.5,-0.3) ;
\draw (3,1.5) [-,color=black, thick]	+(-0.5,-0.5) -- +(-0.5,0.5)-- +(0.5,0.5) -- +(0.5,-0.5)--+(-0.5,-0.5);
\draw (-0.6+3,-0.6+1.5) node {\tiny $+$};
\draw (0.6+3,-0.6+1.5) node {\tiny $+$};
\draw (0.6+3, 0.6+1.5) node {\tiny$-$};
\draw (-0.6+3, 0.6+1.5) node {\tiny$-$};
\draw (3, 0.9+1.5) node {$\omega_5$};

\draw (4.5,1.5) [-,color=black, dotted, directed, rounded corners=7pt]+(-(0.,0.5) --+(0.,0);
\draw (4.5,1.5) [-,color=black, dotted, directed, rounded corners=7pt]+(-(0.,0) --+(0.,-0.5);
\draw (4.5,1.5) [-,color=black,  dotted, directed, rounded corners=7pt]	+(0,0) -- +(0.5,0) ;
\draw (4.5,1.5) [-,color=black,  dotted, directed, rounded corners=7pt]	+(-0.5,0) -- +(0,0) ;
\draw (4.5,1.5) [-,color=black,  thick, rounded corners=7pt]	+(-0.3,-0.5) -- +(-0.3,-0.3) -- +(-0.5,-0.3) ;
\draw (4.5,1.5) [-,color=black, thick]	+(-0.5,-0.5) -- +(-0.5,0.5)-- +(0.5,0.5) -- +(0.5,-0.5)--+(-0.5,-0.5);
\draw (0.95+3,-0.6+1.5) node {\tiny$+$};
\draw (5.15,-0.6+1.5) node {\tiny$-$};
\draw (5.15, 0.6+1.5) node {\tiny$-$};
\draw (3.95, 0.6+1.5) node {\tiny$+$};
\draw (4.5, 0.9+1.5) node {$\omega_7$};

\draw (3,0) [-,color=black, dotted, directed, rounded corners=7pt]+(-(0.,0) --+(0.,0.5);
\draw (3,0) [-,color=black, dotted, directed, rounded corners=7pt]+(-(0.,-0.5) --+(0.,0);
\draw (3,0) [-,color=black,  dotted, directed, rounded corners=7pt]	+(0.5,0) -- +(0,0) ;
\draw (3,0) [-,color=black,  dotted, directed, rounded corners=7pt]	+(0,0) -- +(-0.5,0);
\draw (3,0) [-,color=black,  thick, rounded corners=7pt]	+(-0.3,-0.5) -- +(-0.3,-0.3) -- +(-0.5,-0.3) ;
\draw (3,0) [-,color=black, thick]	+(-0.5,-0.5) -- +(-0.5,0.5)-- +(0.5,0.5) -- +(0.5,-0.5)--+(-0.5,-0.5);
\draw (-0.55+3,-0.6) node {\tiny $-$};
\draw (3.6,-0.6) node {\tiny $-$};
\draw (3.6, 0.6) node {\tiny $+$};
\draw (-0.55+3, 0.6) node {\tiny $+$};
\draw (3, -0.8) node {$\omega_6$};

\draw (4.5,0) [-,color=black, dotted, directed, rounded corners=7pt]+(-(0.,0.5) --+(0.,0);
\draw (4.5,0) [-,color=black, dotted, directed, rounded corners=7pt]+(-(0.,0) --+(0.,-0.5);
\draw (4.5,0) [-,color=black,  dotted, directed, rounded corners=7pt]	+(0,0) -- +(0.5,0) ;
\draw (4.5,0) [-,color=black,  dotted, directed, rounded corners=7pt]	+(-0.5,0) -- +(0,0) ;
\draw (4.5,0) [-,color=black,  thick, rounded corners=7pt]	+(-0.3,-0.5) -- +(-0.3,-0.3) -- +(-0.5,-0.3) ;
\draw (4.5,0) [-,color=black, thick]	+(-0.5,-0.5) -- +(-0.5,0.5)-- +(0.5,0.5) -- +(0.5,-0.5)--+(-0.5,-0.5);
\draw (3.95,-0.6) node {\tiny$-$};
\draw (5.15,-0.6) node {\tiny$+$};
\draw (5.15, 0.6) node {\tiny$+$};
\draw (3.95, 0.6) node {\tiny$-$};
\draw (4.5, -0.8) node {$\omega_8$};


\draw (6,1.5) [-,color=black, dotted, directed, rounded corners=7pt]+(-(0.,0.5) --+(0.,0);
\draw (6,1.5) [-,color=black, dotted, directed, rounded corners=7pt]+(-(0.,-0.5) --+(0.,0.);
\draw (6,1.5) [-,color=black,  dotted, directed, rounded corners=7pt]	+(0,0) -- +(0.5,0) ;
\draw (6,1.5) [-,color=black,  dotted, directed, rounded corners=7pt]	+(0,0) -- +(-0.5,0) ;
\draw (6,1.5) [-,color=black,  thick, rounded corners=7pt]	+(-0.3,-0.5) -- +(-0.3,-0.3) -- +(-0.5,-0.3) ;
\draw (6,1.5) [-,color=black, thick]	+(-0.5,-0.5) -- +(-0.5,0.5)-- +(0.5,0.5) -- +(0.5,-0.5)--+(-0.5,-0.5);
\draw (-0.6+6,-0.6+1.5) node {\tiny $+$};
\draw (0.6+6,-0.6+1.5) node {\tiny $+$};
\draw (0.6+6, 0.6+1.5) node {\tiny$+$};
\draw (-0.6+6, 0.6+1.5) node {\tiny$-$};
\draw (6, 0.9+1.5) node {$\omega_9$};

\draw (7.5,1.5) [-,color=black, dotted, directed, rounded corners=7pt]+(-(0.,0) --+(0.,0.5);
\draw (7.5,1.5) [-,color=black, dotted, directed, rounded corners=7pt]+(-(0.,0) --+(0.,-0.5);
\draw (7.5,1.5) [-,color=black,  dotted, directed, rounded corners=7pt]	+(0.5,0) -- +(0,0) ;
\draw (7.5,1.5) [-,color=black,  dotted, directed, rounded corners=7pt]	+(-0.5,0) -- +(0,0) ;
\draw (7.5,1.5) [-,color=black,  thick, rounded corners=7pt]	+(-0.3,-0.5) -- +(-0.3,-0.3) -- +(-0.5,-0.3) ;
\draw (7.5,1.5) [-,color=black, thick]	+(-0.5,-0.5) -- +(-0.5,0.5)-- +(0.5,0.5) -- +(0.5,-0.5)--+(-0.5,-0.5);
\draw (0.95+6,-0.6+1.5) node {\tiny$+$};
\draw (8.15,-0.6+1.5) node {\tiny$-$};
\draw (8.15, 0.6+1.5) node {\tiny$+$};
\draw (6.95, 0.6+1.5) node {\tiny$+$};
\draw (7.5, 0.9+1.5) node {$\omega_{11}$};

\draw (6,0) [-,color=black, dotted, directed, rounded corners=7pt]+(-(0.,0.5) --+(0.,0);
\draw (6,0) [-,color=black, dotted, directed, rounded corners=7pt]+(-(0.,-0.5) --+(0.,0);
\draw (6,0) [-,color=black,  dotted, directed, rounded corners=7pt]	+(0,0) -- +(0.5,0) ;
\draw (6,0) [-,color=black,  dotted, directed, rounded corners=7pt]	+(0,0) -- +(-0.5,0);
\draw (6,0) [-,color=black,  thick, rounded corners=7pt]	+(-0.3,-0.5) -- +(-0.3,-0.3) -- +(-0.5,-0.3) ;
\draw (6,0) [-,color=black, thick]	+(-0.5,-0.5) -- +(-0.5,0.5)-- +(0.5,0.5) -- +(0.5,-0.5)--+(-0.5,-0.5);
\draw (-0.55+6,-0.6) node {\tiny $-$};
\draw (6.6,-0.6) node {\tiny $-$};
\draw (6.6, 0.6) node {\tiny $-$};
\draw (-0.55+6, 0.6) node {\tiny $+$};
\draw (6, -0.8) node {$\omega_{10}$};

\draw (7.5,0) [-,color=black, dotted, directed, rounded corners=7pt]+(-(0.,0) --+(0.,0.5);
\draw (7.5,0) [-,color=black, dotted, directed, rounded corners=7pt]+(-(0.,0) --+(0.,-0.5);
\draw (7.5,0) [-,color=black,  dotted, directed, rounded corners=7pt]	+(0.5,0) -- +(0,0) ;
\draw (7.5,0) [-,color=black,  dotted, directed, rounded corners=7pt]	+(-0.5,0) -- +(0,0) ;
\draw (7.5,0) [-,color=black,  thick, rounded corners=7pt]	+(-0.3,-0.5) -- +(-0.3,-0.3) -- +(-0.5,-0.3) ;
\draw (7.5,0) [-,color=black, thick]	+(-0.5,-0.5) -- +(-0.5,0.5)-- +(0.5,0.5) -- +(0.5,-0.5)--+(-0.5,-0.5);
\draw (6.95,-0.6) node {\tiny$-$};
\draw (8.15,-0.6) node {\tiny$+$};
\draw (8.15, 0.6) node {\tiny$-$};
\draw (6.95, 0.6) node {\tiny$-$};
\draw (7.5, -0.8) node {$\omega_{12}$};


\draw (1.5,-1.5) [-,color=black, dotted, directed, rounded corners=7pt]+(-(0.,0.5) --+(0.,0);
\draw (1.5,-1.5) [-,color=black, dotted, directed, rounded corners=7pt]+(-(0.,-0.5) --+(0.,0.);
\draw (1.5,-1.5) [-,color=black,  dotted, directed, rounded corners=7pt]	+(0.5,0) -- +(0,0) ;
\draw (1.5,-1.5) [-,color=black,  dotted, directed, rounded corners=7pt]	+(-0.5,0) -- +(0,0) ;
\draw (1.5,-1.5) [-,color=black,  thick, rounded corners=7pt]	+(-0.3,-0.5) -- +(-0.3,-0.3) -- +(-0.5,-0.3) ;
\draw (1.5,-1.5) [-,color=black, thick]	+(-0.5,-0.5) -- +(-0.5,0.5)-- +(0.5,0.5) -- +(0.5,-0.5)--+(-0.5,-0.5);
\draw (0.95,-0.6-1.5) node {\tiny $+$};
\draw (0.6+1.5,-0.6-1.5) node {\tiny $+$};
\draw (0.6+1.5, 0.6-1.5) node {\tiny$-$};
\draw (0.95, 0.6-1.5) node {\tiny$+$};
\draw (1.5, -2.25) node {$\omega_{13}$};

\draw (3.,-1.5) [-,color=black, dotted, directed, rounded corners=7pt]+(-(0.,0.5) --+(0.,0);
\draw (3,-1.5) [-,color=black, dotted, directed, rounded corners=7pt]+(-(0.,-0.5) --+(0.,0);
\draw (3,-1.5) [-,color=black,  dotted, directed, rounded corners=7pt]	+(0.5,0) -- +(0,0) ;
\draw (3,-1.5) [-,color=black,  dotted, directed, rounded corners=7pt]	+(-0.5,0) -- +(0,0) ;
\draw (3,-1.5) [-,color=black,  thick, rounded corners=7pt]	+(-0.3,-0.5) -- +(-0.3,-0.3) -- +(-0.5,-0.3) ;
\draw (3,-1.5) [-,color=black, thick]	+(-0.5,-0.5) -- +(-0.5,0.5)-- +(0.5,0.5) -- +(0.5,-0.5)--+(-0.5,-0.5);
\draw (0.65+3,-0.6-1.5) node {\tiny$-$};
\draw (2.45,-0.6-1.5) node {\tiny$-$};
\draw (2.45, 0.6-1.5) node {\tiny$-$};
\draw (3.65, 0.6-1.5) node {\tiny$+$};
\draw (3, -2.25) node {$\omega_{14}$};

\draw (4.5,-1.5) [-,color=black, dotted, directed, rounded corners=7pt]+(-(0.,0) --+(0.,0.5);
\draw (4.5,-1.5) [-,color=black, dotted, directed, rounded corners=7pt]+(-(0.,0) --+(0.,-0.5);
\draw (4.5,-1.5) [-,color=black,  dotted, directed, rounded corners=7pt]	+(0,0) -- +(0.5,0) ;
\draw (4.5,-1.5) [-,color=black,  dotted, directed, rounded corners=7pt]	+(0,0) -- +(-0.5,0);
\draw (4.5,-1.5) [-,color=black,  thick, rounded corners=7pt]	+(-0.3,-0.5) -- +(-0.3,-0.3) -- +(-0.5,-0.3) ;
\draw (4.5,-1.5) [-,color=black, thick]	+(-0.5,-0.5) -- +(-0.5,0.5)-- +(0.5,0.5) -- +(0.5,-0.5)--+(-0.5,-0.5);
\draw (3.95,-0.6-1.5) node {\tiny $-$};
\draw (5.1,-0.6-1.5) node {\tiny $+$};
\draw (5.1, 0.6-1.5) node {\tiny $+$};
\draw (-0.55+4.5, 0.6-1.5) node {\tiny $+$};
\draw (4.5, -2.25) node {$\omega_{15}$};

\draw (6,-1.5) [-,color=black, dotted, directed, rounded corners=7pt]+(-(0.,0) --+(0.,0.5);
\draw (6,-1.5) [-,color=black, dotted, directed, rounded corners=7pt]+(-(0.,0) --+(0.,-0.5);
\draw (6,-1.5) [-,color=black,  dotted, directed, rounded corners=7pt]	+(0,0) -- +(0.5,0) ;
\draw (6,-1.5) [-,color=black,  dotted, directed, rounded corners=7pt]	+(0,0) -- +(-0.5,0) ;
\draw (6,-1.5) [-,color=black,  thick, rounded corners=7pt]	+(-0.3,-0.5) -- +(-0.3,-0.3) -- +(-0.5,-0.3) ;
\draw (6,-1.5) [-,color=black, thick]	+(-0.5,-0.5) -- +(-0.5,0.5)-- +(0.5,0.5) -- +(0.5,-0.5)--+(-0.5,-0.5);
\draw (5.45,-0.6-1.5) node {\tiny$+$};
\draw (6.65,-0.6-1.5) node {\tiny$-$};
\draw (6.65, 0.6-1.5) node {\tiny$-$};
\draw (5.45, 0.6-1.5) node {\tiny$-$};
\draw (6, -0.8-1.5) node {$\omega_{16}$};

\end{tikzpicture}
\end{center}
\end{minipage}
\caption{Boltzmann weights of the face version of the eight-vertex model. The face weight is obtained from the allowed configuration for the eight-vertex model, which are indicated by the dotted oriented lines at the center of the face weight.}
\label{face6V}
\end{figure}

The face version of the eight-vertex model is depicted in the Figure \ref{face6V}. The face model is considered to be made of two copies of the eight-vertex model, whose spin/height are assigned as in Figure \ref{face6V}.

For instance, the face weights related to the vertex model weight $\mathfrak{a}(\lambda)$ are given by
\bear
\omega_1=\weight[+][+][+][+][\lambda], \quad \omega_3=\weight[-][+][-][+][\lambda],\nonumber \\
\omega_2=\weight[-][-][-][-][\lambda], \quad \omega_4=\weight[+][-][+][-][\lambda].
\ear

In order to have the Yang-Baxter equation naturally fulfilled, we suitably chose the face weights $\omega_i$ to correspond to the statistical weights of the eight-vertex model denoted as $\mathfrak{a}(\lambda)$ , $\mathfrak{b}(\lambda)$, $\mathfrak{c}(\lambda)$ and $\mathfrak{d}(\lambda)$, such that
\begin{alignat}{5}
\omega_1 &=\omega_2 &=\omega_3 &=\omega_4 &=\mathfrak{a}(\lambda), \nonumber \\
\omega_5 &=\omega_6 &=\omega_7 &=\omega_8 &=\mathfrak{b}(\lambda),  \\
\omega_9 &=\omega_{10} &=\omega_{11} &=\omega_{12} &=\mathfrak{c}(\lambda), \nonumber \\
\omega_{13} &=\omega_{14} &=\omega_{15} &=\omega_{16} &=\mathfrak{d}(\lambda). \nonumber
\end{alignat}

Taking the logarithmic derivative of the IRF transfer matrix associated with the eight-vertex model, we obtain an one-dimensional spin chain with interaction of three spins,
\bear
{\cal H}^{\text{8v}}_{\text{IRF}}=
\sum_{i=1}^L  (\Gamma+1)\sigma_i^x +	(\Gamma-1)\sigma_{i-1}^z \sigma_i^x\sigma_{i+1}^z+\Delta(\Gamma+1)\left(\sigma_{i-1}^z \sigma_{i+1}^z-1\right),
\label{Hirf8v}
\ear
where $\sigma^{\alpha}$ for $\alpha=x,y,z$ are the standard Pauli matrices and the anistropy parameters are given by $\Delta=\frac{\mathfrak{a}^2+\mathfrak{b}^2-\mathfrak{c}^2-\mathfrak{d}^2}{2(\mathfrak{a}\mathfrak{b}+\mathfrak{c}\mathfrak{d})}$ and $\Gamma=\tfrac{\mathfrak{c}\mathfrak{d}}{\mathfrak{a}\mathfrak{b}}$.

From now on, we specialize the Hamiltonian (\ref{Hirf8v}) for $\Gamma=\mathfrak{d}=0$, so it becomes the face version of the six-vertex model simply denoted as,
\bear
{\cal H}_{\text{IRF}}=
\sum_{i=1}^L  \sigma_i^x -\sigma_{i-1}^z\sigma_i^x \sigma_{i+1}^z + \Delta\left(\sigma_{i-1}^z \sigma_{i+1}^z-1\right),
\label{Hirf}
\ear
which has a continuous $U(1)$ symmetry and a discrete $\mathbb{Z}_2$ symmetry. This is due to the fact that for,
\eq
\Sigma^z= \sum_{j=1}^L \sigma_j^z \sigma_{j+1}^z,\qquad \Pi^x=\prod_{j=1}^L \sigma_j^x, \qquad \left[\Pi^x,\Sigma^z\right]=0,
\en
one has that
\begin{alignat}{2}
	\left[H_{\text{IRF}},\Sigma^z\right]&=0, \qquad &\left[H_{\text{IRF}},\Pi^x\right]&=0, \nonumber\\
	\left[T_{\text{IRF}}(\lambda),\Sigma^z\right]&=0, \qquad &\left[T_{\text{IRF}}(\lambda),\Pi^x\right]&=0.
\end{alignat}
In addition, the Hamiltonian (\ref{Hirf}) has $\mathbb{Z}_2\times \mathbb{Z}_2$ for even $L$, since $[{\cal H}_{\text{IRF}},\Pi^x_{\text{even}}]=[{\cal H}_{\text{IRF}},\Pi^x_{\text{odd}}]=0$.

The parameter $\Delta$ in (\ref{Hirf}) is precisely the anisotropy parameter of the Heisenberg (XXZ) model, whose Hamiltonian can be conveniently written as,
\eq
{\cal H}_{\text{XXZ}}(\phi)= \sum_{i=1}^{L-1}   \sigma_i^x \sigma_{i+1}^x +\sigma_i^y \sigma_{i+1}^y+ \Delta\left(\sigma_{i}^z \sigma_{i+1}^z-1\right) + h_{L,1},
\label{Hxxz}
\en
where boundary term $h_{L,1}={\cal G}_L^{-1} \cdot \left(\sigma_L^x \sigma_{1}^x +\sigma_L^y \sigma_{1}^y+ \Delta\left(\sigma_{L}^z \sigma_{1}^z-1\right) \right) \cdot {\cal G}_L$ is given in terms of the twist angle $\phi$ through the diagonal matrix ${\cal G}\!=\!\diag(\exp(2\im  \phi),1)$.

The Hamiltonian (\ref{Hxxz}) is obtained from the logarithmic derivative of the six-vertex transfer matrix with twisted boundary condition,
\eq
T_{\text{6v}}(\lambda;\phi)=\tr_{\cal A}\left[ {\cal G}_{\cal A} R_{{\cal A}L}(\lambda) \dots R_{{\cal A}2}(\lambda)R_{{\cal A}1}(\lambda) \right],
\en
where the $R$-matrix
\eq
R(\lambda)=\left(\begin{matrix}
	\mathfrak{a}(\lambda) & 0 & 0 & 0 \\
	0 & \mathfrak{b}(\lambda) & \mathfrak{c}(\lambda) & 0 \\
	0 & \mathfrak{c}(\lambda) & \mathfrak{b}(\lambda) & 0 \\
	0 & 0 & 0 & \mathfrak{a}(\lambda) \\
\end{matrix}\right),
\en
and the vertex weights $\mathfrak{a}(\lambda)$ , $\mathfrak{b}(\lambda)$ and $\mathfrak{c}(\lambda)$ are given such that the Yang-Baxter equation is satisfied \cite{BAXTER,BOOK-KBI}.

The mapping of between the IRF partition function $Z_{\text{IRF}}$ and the six-vertex model partition function $Z_{\text{6v}}(\phi)$ under special boundary conditions\cite{BAXTER,KLUMPER93}
implies that the spectrum of the both transfer matrices $T_{\text{IRF}}(\lambda)$ and $T_{\text{6v}}(\lambda;\phi)$ are related. The same applies for the associated Hamiltonians $\cal{H}_{\text{IRF}}$ and $\cal{H}_{\text{XXZ}}(\phi)$.

More explicitly, the transfer matrix $T_{\text{IRF}}(\lambda)$ with periodic boundary is related with $T_{\text{6v}}(\lambda;\phi)$ with twisted angles $\phi=0$ and $\phi=\pi/2$, such that
\eq
U T_{\text{IRF}}(\lambda) U^{t} = T_{\text{6v}}^{\text{even}}(\lambda;0) \oplus T_{\text{6v}}^{\text{even}}(\lambda;\pi/2), \label{VertexToIRF}
\en
in the sector of even number of spin flips. $U$ is the matrix that diagonalizes the operator $\Pi^x$, given by
\eq
U=\frac{1}{\sqrt2}\left(\begin{array}{cc}
	I & \Pi^x_{L-1} \\
	\Pi^x_{L-1} & -I
\end{array}\right),
\en
where $I$ is $2^{L-1}\times 2^{L-1}$ identity matrix and  $\Pi^x_{L-1}=\prod_{j=1}^{L-1}\sigma^x_j$. Therefore, the $\phi=0$ sector is associated to the  positive parity eigenvalues and $\phi=\pi/2$ to the negative ones.
Naturally, the same applies for the Hamiltonian,
\eq
U {\cal H}_{\text{IRF}} U^{t} = {\cal H}_{\text{XXZ}}^{\text{even}}(0) \oplus {\cal H}_{\text{XXZ}}^{\text{even}}(\pi/2).
\en

Therefore, the spectrum of the IRF transfer matrix can be obtained from the spectrum of the six-vertex transfer matrix under special boundaries and in the even spin flip sectors.
For completeness, let us mention that there exists a similar relation between the IRF transfer matrix with antiperiodic boundary condition and the six-vertex transfer matrix with twisted angles $\phi=0$ and $\phi=\pi/2$ in the sector of odd spin flips\cite{KLUMPER93}. Here we find that the IRF Hamiltonian and the $U(1)$ charge boundary operators are modified according to $\sigma_{0}^z = - \sigma_{L}^{z} $ and $\sigma_{L+1}^{z} = -\sigma_{1}^{z}$. In terms of two-point long-distance correlations, this implies an additional contribution of $\pi$ on the momentum governing the oscillatory behavior of the correlation function.

It is worth noticing that the Hamiltonian (\ref{Hirf8v}) appeared previously in the context of the bond-site transformation \cite{POZSGAY} and its limiting case (\ref{Hirf}) has already appeared in the context of Clifford group transformations \cite{H6v}. Besides that, a Hamiltonian similar to (\ref{Hirf}) has also just appeared in the context of chromatic algebras in an interesting recent work \cite{FENDLEY} where it was termed as Ising zigzag ladder (${\cal H}_{\text{zig}}$). Although both models are alike, they have different symmetries for general values of the parameter $\Delta$, since $[{\cal H}_{\text{zig}},\Sigma^z]=0$ only for $\Delta=1$. This point is where  both models can be related, since ${\cal H}_{\text{IRF}}(\Delta=-1)=-2 {\cal H}_{\text{zig}}(\Delta=1)+2 $. Besides, the three-spins interaction terms as in (\ref{Hirf}) are important on their own right, since they arise in the context of symmetry-protected topological phase \cite{NATURE}.

\section{QTM approach and non-linear integral equations}\label{NLIE}

The $U(1)$ symmetry of the model, generated by the $\Sigma^z$, can be explored to include an additional term to the Hamiltonian as a kind of Ising interaction. This amounts to the inclusion of a horizontal seam in the Trotter-Suzuki decomposition, which in terms of the quantum transfer matrix becomes simply a twist. Therefore, the thermodynamical properties are obtained through the free-energy as \cite{KLUMPER92},
\eq
f=-\frac{1}{\beta} \lim_{L\rightarrow \infty} \frac{\ln{Z_L}}{L} = -\frac{1}{\beta} \lim_{N\rightarrow \infty} \ln\Lambda_{\text{max}}, \label{Goal}
\en
where $Z_L=\mbox{Tr}\left[\exp{\left(-\beta H_{\text{IRF}} -\beta J\Sigma^z\right)}\right]$ is the
partition function and $J$ is an Ising like coupling.
In the last equality of (\ref{Goal}), the thermodynamical limit has already been taken. Therefore, to obtain finite temperature properties, we only need the largest eigenvalue of the quantum transfer matrix ($\Lambda_{\text{max}}$) in the large Trotter number limit.

The relation (\ref{VertexToIRF}) also applies for the quantum transfer matrix, so we may write the quantum transfer matrix eigenvalues as,
\eq
\Lambda(x)=\begin{cases}
	\Lambda_{\text{6v}}^{\text{even}}(x;\phi=0), \\
	\Lambda_{\text{6v}}^{\text{even}}(x;\phi=\pi/2),
\end{cases}\label{MAP}
\en
which results in the eigenvalue expression \cite{BAXTER,LIEB},
\begin{equation}
\Lambda(x)= \lambda_1(x)+\lambda_2(x),  \label{eigenvalue1}
\end{equation}
and the Bethe ansatz equations,
\begin{equation}
	\frac{\lambda_1(x_j)}{\lambda_2(x_j)}=-1, \qquad j=1,\ldots,n, \label{BA}
\end{equation}
where
\begin{eqnarray}
  \lambda_1(x) &=& {\rm e}^{-\beta J+2 \im \phi}(\mathfrak{a}(\im x+\lambda))^{\frac{N}{2}}(\mathfrak{b}(-\im x+\lambda))^{\frac{N}{2}} \prod_{j=1}^{n} \frac{\mathfrak{a}(\im x_j-\im x)}{\mathfrak{b}(\im x_j-\im x)}, \nonumber\\
  \lambda_2(x) &=&
 {\rm e}^{\beta J}(\mathfrak{b}(\im x+\lambda))^{\frac{N}{2}}(\mathfrak{a}(-\im x+\lambda))^{\frac{N}{2}} \prod_{j=1}^{n} \frac{\mathfrak{a}(\im x-\im x_j)}{\mathfrak{b}(\im x-\im x_j)}, \label{eigenvalue2}
\end{eqnarray}
where $n$ corresponds to the $U(1)$ sector\footnote{The description of the $U(1)$ symmetry inherited by the quantum transfer matrix can be realized either in terms of the number of down spins or the staggered number of down spins. In the present formulation of the Trotter-Suzuki decomposition it is the number of down spins.} and is taken as even number and $\phi=0,~\frac{\pi}{2}$ to establish the connection with the IRF transfer matrix eigenvalues.

The leading eigenvalue of the IRF quantum transfer matrix $\Lambda_{\text{max}}=\Lambda(0)$ is obtained for $\phi=0$, which is the same as the six-vertex quantum transfer matrix leading eigenvalue. Therefore, one can use exactly the same non-linear integral equations \cite{KLUMPER92} to describe the thermodynamical properties of the model and the ground state phase diagram.

In order to exemplify, we define the Boltzmann weights for the  critical regime $|\Delta| < 1$, which can be parameterized as $\mathfrak{a}(\lambda)=1$, $\mathfrak{b}(\lambda)=\sin(\lambda)/\sin(\lambda+\gamma)$ and $\mathfrak{c}(\lambda)=\sin(\gamma)/\sin(\lambda+\gamma)$, which implies that $\Delta=\cos{\gamma}$ and $0<\gamma<\pi$. Similarly can be done for $|\Delta|>1$ \cite{BAXTER}.

The free-energy (\ref{Goal}) can be obtained from,
\eq
\ln \Lambda(x)= -\beta e_0(x) + \im \phi+ \left( K \ast \ln{B \bar{B}}\right)(x), \label{NLIEeigenvalue}
\en
where the ground state energy $e_0(0)$ is obtained from the function,
\eq
e_0(x)=\begin{cases}
\displaystyle	 -2 \sin \gamma \int_{-\infty}^{\infty}{\rm e}^{\im k x} \frac{\sinh \frac{k}{2} (\pi-\gamma)}{2 \sinh\frac{k\pi}{2} \cosh \frac{k \gamma}{2}} {\rm d}k, & 0< \Delta \leq 1\\
\displaystyle  -2  \sinh{\gamma}\sum_{k=-\infty}^{\infty}
{\rm  e}^{\im 2 k x}\frac{{\rm  e}^{-\gamma |k|}}{\cosh{\gamma k}}, & \Delta > 1,
       \end{cases}
\en
and
\eq
 K(x)=\begin{cases}
	\frac{\pi}{\cosh{ \pi x}}, & 0< \Delta \leq 1,  \\
	\displaystyle\sum_{k=-\infty}^{\infty}\tfrac{e^{\im 2 k x}}{2\cosh{\gamma k}}, & \Delta>1.
\end{cases}
\en
The symbol $\ast$ denotes convolution $f*g(x)=\frac{1}{2 \pi}\int_{-a}^{a} f(x-y)g(y)dy$ where $a\rightarrow \infty$ for $0< \Delta \leq 1$ and $a=\pi/2$
for $\Delta>1$.

The auxiliary functions $b(x)$, $\bar{b}(x)$ and its simply related functions
$B(x)=b(x)+1$ and $\bar{B}(x)=\bar{b}(x)+1$ are solutions of the following set
of non-linear integral equations \cite{KLUMPER92},
\bear
	\ln{b(x)}&=&  d_{+}(x)  +  \left( \!F\!\ast\! \ln{B}\right)\!(x) -
	\left(\! F\!\ast\! \ln{\bar{B}}\right)\!(x+\im\gamma), \label{NLIE1}\\
	\ln{\bar{b}(x)}&=& d_{-}(x)  - \left(\! F \!\ast\! \ln{B}\right)\!(x-\im\gamma) + \left(\! F\!\ast\! \ln{\bar{B}}\right)\!(x).	
	\label{NLIE2}
\ear
The driving term $d_{\pm}(x)$ is given by
\eq
d_{\pm}(x)=\begin{cases}
	\displaystyle	-2 \beta \frac{\sin{\gamma}}{\gamma} \frac{\pi}{\cosh{ (\pi x/\gamma) }} \pm  (\beta J-\im\phi) \frac{\pi}{\pi-\gamma}, & 0< \Delta \leq 1,  \\
	\displaystyle -2 \beta \sinh{\gamma} \sum_{k=-\infty}^{\infty}
	\frac{e^{\im 2 k x}}{\cosh{\gamma k}} \pm  (\beta J-\im\phi), & \Delta>1,
\end{cases}
\en
and the Kernel function is given by
\eq
F(x)=\begin{cases}
	\displaystyle\int_{-\infty}^{\infty}\frac{\sinh{\frac{k}{2}(\pi-2 \gamma)
				} ~e^{\im k x}}{2\sinh{\frac{k}{2}(\pi-\gamma)
				}\cosh{\frac{k\gamma}{2} }}{\rm d}k, & 0< \Delta \leq 1, \\
	\displaystyle 	\sum_{k=-\infty}^{\infty} \frac{e^{-\gamma |k|}}{\cosh{ k \gamma}} e^{\im 2 k x}, 	& \Delta > 1.
\end{cases}
\en

The above non-linear equations are a consequence of the analyticity properties that the Bethe ansatz equations (\ref{BA}) imply. For the model (\ref{Hirf}), this hypothesis is not only true for the largest eigenvalue, but also for the first few sub-leading eigenvalues. This means that not only the leading eigenvalue of the IRF quantum transfer matrix is free of zeros inside the analiticity strip, but this also applies for the first few subsequent ones as will be discussed later.

\subsection{Low-temperature behavior }

The above non-linear integral equations are well suited to the evaluation of the universal part of the spectrum. Here we reframe the di-log trick \cite{KLUMPER93} within our conventions to derive results for the low-temperature behavior and $J$ dependence of the largest eigenvalues for $|\Delta|<1$.

In order to do that, we first integrate the Eqs.(\ref{NLIE1}-\ref{NLIE2}) by parts, which can be written as,
\eq
\ln \frac{b_i(x)}{b_i(\infty)} = d_{i}(x)+ \!F_{i j}^{\text{int}}\!\ast\! \ln'{B_j},
\en
where $b_1(x)=b(x)$, $b_2(x)=\bar{b}(x)$ and $d_1(x)=d_+(x)$, $d_2(x)=d_-(x)$ are two-component vectors and $\!F_{i j}^{\text{int}}$ is a two by two tensor whose components may be inferred from (\ref{NLIE1}-\ref{NLIE2}). Here we are summing over repeated indices. Moreover, $F_{ij}^{\text{int}}(x)$ is obtained from the integration of $F_{ij}(x)$.

Similarly, we rewrite the eigenvalue expression as,
\eq
\ln \frac{\Lambda(x)}{\Lambda(\infty)}= -\beta e_0(x) + \im \phi + \frac{1}{2 \pi}\int_{-\infty}^{\infty} D_j(x-s) \ln' B_j(s) {\rm d }s,
\en
where we have defined the integrated driving-term,
\eq
D_j(x)= \frac{1}{\im} \ln \left[\frac{{\rm e}^{\frac{\pi x}{\gamma}}-\im}{{\rm e}^{\frac{\pi x}{\gamma}}+\im}\right].
\en
To compute the low temperature behavior it is necessary to expand the dependence of the integral,
\eq
G_{\pm}=\pm \frac{1}{2 \pi}\int_{0}^{\pm \infty} D_j(x-s) \ln' B_j(s){\rm d}s,
\en
in powers of $\beta$. This can be done through the change of variables $s \rightarrow s\pm\frac{\gamma}{\pi}\ln \beta$:
\begin{multline}
G_+= \frac{1}{2} \ln B(0)\bar{B}(0) -\frac{1}{2} \ln B(\infty)\bar{B}(\infty)+\\\sum_{k=0}^{\infty} \frac{{\rm e}^{\frac{\pi x }{\gamma}(2 k+1)}{(-1)}^k}{\pi (2k +1) \beta^{2 k+1}} \underbrace{\int_{-\frac{\gamma}{\pi} \ln \beta}^{\infty} {\rm e}^{-\frac{\pi  s}{\gamma} (2 k+1)} \ln'(B_+(s) \bar{B}_+(s)){\rm d}s}_{I_+^{(k)}},
\label{gplus}
\end{multline}
\eq
G_-= -\sum_{k=0}^{\infty} \frac{{\rm e}^{-\frac{\pi x }{\gamma}(2 k+1)}{(-1)}^k}{\pi (2k +1) \beta^{2 k+1}} \underbrace{\int_{-\infty}^{\frac{\gamma}{\pi} \ln \beta} {\rm e}^{\frac{\pi  s}{\gamma} (2 k+1)} \ln'(B_-(s) \bar{B}_-(s)){\rm d}s }_{I_-^{(k)}},
\label{gminus}
\en
so the non-trivial part comes from the integrals $I_{\pm}^{(k)}$. To obtain these, we take the scaling limit in the non-linear integral equations by setting $b_{j\pm}(x)=b_{j}(x\pm \frac{\gamma}{\pi} \ln \beta)$ and construct the desired expressions. For instance, we may realize an inner product of the scaled equations with the vector $\ln' B_+(x)$, to write $I_+^{(0)}$ in the form,
\eq
\!\!\int_{-\infty}^{\infty}\!\!\!\!\!\! \ln \tfrac{b_{j+}(s)}{b_j(\infty)} \ln' \! B_{j+}(s)  {\rm d}s - \tfrac{1}{2 \pi}\!\!\iint_{-\infty}^{\infty} \!\!\!\!\!\! F_{i j}^{\text{int}}(x-s)\ln'\! B_{i+}(x)  \ln'\! B_{j+}(s)  {\rm d }s{\rm d }x \!=\! - 4 \tfrac{\sin \gamma}{\gamma} \pi I_+^{(0)}. \label{Iplus}
\en
We apply the di-log identity to the first term on the left hand side,
\eq
\int_{0}^{\infty}\!\!\!  \ln \tfrac{b_i(s)}{b_i(\infty)} \!\ln' \!B_i(s) {\rm d}s \!=\! \sum_i \!\! \int_{b_i(0) \rightarrow 0}^{b_i(\infty)}\!\!\! \tfrac{\ln z}{1+z}{\rm d}z -\ln B_i(\infty) \ln b_i(\infty) \!=\! - \frac{\pi^2}{6}-\frac{1}{2}\!\ln^2 \!b(\infty).
\en
Notice that we have identified $b_{i+}(-\infty) \approx b_{i}(0)$.

The second term on the left hand side (\ref{Iplus}) can be evaluated because of the Kernel symmetry: $F_{i j}^{\text{int}} (x-s)+F_{j i}^{\text{int}} (s-x)= C$. The constant $C$ depends on the integration limit which defines $F_{i j}^{\text{int}} (x)$. We took $F_{i j}^{\text{int}} (\infty)=0$, so we find $C=-\pi \hat{F}(0)=-\pi \frac{\pi-2 \gamma}{2 (\pi -\gamma)}$. Therefore we get
\eq
\frac{1}{2 \pi}\int_{-\infty}^{\infty} \int_{-\infty}^{\infty} F_{i j}^{\text{int}} (x-s)\ln' B_{i+}(x)  \ln' B_{j+}(s)  {\rm d }s{\rm d }x = -\frac{1}{2} \hat{F}(0) \ln^2 b(\infty),
\en
and thus we can evaluate $I_+^{(0)}$ from (\ref{Iplus}). Similar considerations apply to the evaluation of the $k=0$ correction for the $G_-$. By considering only the $k=0$ corrections in (\ref{gplus}-\ref{gminus}), we obtain,
\begin{equation}
\ln \Lambda(x)+\beta e_0(x)=\im \phi +  \frac{2\cosh \frac{\pi x}{\gamma}}{\beta} \frac{\gamma}{\sin \gamma}\left[\frac{1}{24}+\frac{{ \left(\frac{\beta J-\im \phi}{\pi}\right)}^2}{4 (1-\frac{\gamma}{\pi})}\right]. \label{k0correction}
\end{equation}
At $J=0$ this is sufficient to determine the universal data which, in the present case, gives the linear dependence of the specific heat.
For  $J\neq0$, the above calculation does not take into account all the $O(\frac{1}{\beta})$ corrections due to $\beta$ dependent asymptotic limit of the auxiliary functions. On the other hand, the imaginary part of the $O(1)$ correction is already given by (\ref{k0correction}), as discussed in the next section.

\section{Phase diagram and long distance correlation}\label{res}

The mapping between vertex and IRF models (\ref{VertexToIRF}) implies that the thermodynamical properties of the ${\cal H}_{IRF}$ are the same as of the $XXZ$ spin chain. This is due to the fact that the leading eigenvalue of the IRF quantum transfer matrix is precisely the leading eigenvalue of the $XXZ$ quantum transfer matrix at even sector with periodic boundary condition ($\phi=0$). Therefore the thermodynamical properties (like free-energy, entropy and others) and the ground state phase diagram are exactly the same for both spin chains.

\begin{figure}[h]
\begin{center}
	\includegraphics[width=0.5\linewidth, angle=-90]{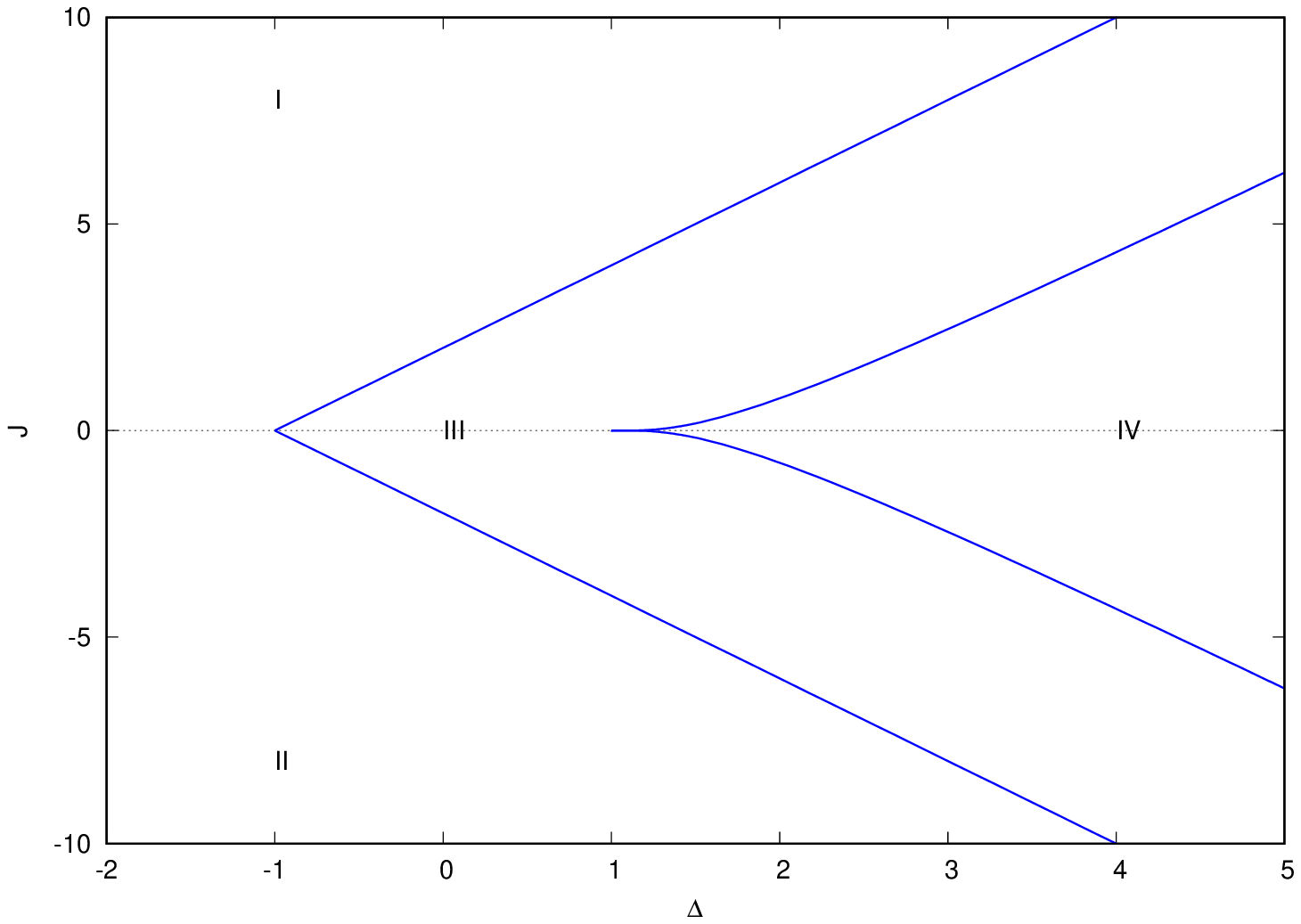}
		\end{center}
	\caption{Phase diagram of the spin chain. The phase I (N\'eel antiferromagnetic), II (ferromagnetic) and IV (dimerized antiferromagnetic) are gapped, while the phase III (dimerized antiferromagnetic) is gapless. }
	\label{phase-diagram}
\end{figure}

Nevertheless, the first sub-dominant eigenvalues of the IRF quantum transfer matrix comes from the twisted part of the spectrum with $\phi=\tfrac{\pi}{2}$, which implies that the correlation length of the ${\cal H}_{IRF}$ differ from the value of the $XXZ$ spin chain with periodic boundary condition. Therefore, the predominant order in each phase is manifestly different, see Figure \ref{phase-diagram}.

For simplicity, let us first discuss the gapped phases where one can inspect the exact ordering through the large coupling constant limits. Phase $\textrm{I}$ (see Figure \ref{phase-diagram}) can be analyzed in terms of a large and negative anisotropy parameter $\Delta$ with a residual positive Ising coupling $J>0$. In this phase, the next-to-nearest neighbors spins align in parallel manner, whereas for positive $J$ the nearest neighbors are aligned anti-parallelly. This leave us with two such states, both of them with N\'eel long-range order $\ket{\dots\uparrow\downarrow\uparrow\downarrow\uparrow\dots}$. One could use the $\Pi^x$ symmetry and verify that under spin-reversal one can build two states with eigenvalues $\pm1$. On the other hand, by considering negative $J$, we obtain full alignment among all spins. This corresponds to the ferromagnetic phase $\textrm{II}$ with long-range order $\ket{\dots\uparrow\uparrow\uparrow\uparrow\uparrow\dots}$.

Nevertheless, by assuming e.g positive $\Delta$ and relatively small $J$, the chain will decompose into two interpenetrating sub-lattices, where each of them has, individually, a N\'eel order. Independently of the sign of $J$, in the phase IV the spin chain will be frustrated. Degeneracy of the ground-state depends on the remainder of the division of the number of sites by 4. In any case, however, the ground-state is a linear combination of states that have a dimerized antiferromagnetic order.

The critical region III, in between phases \textrm{I,~II} and \textrm{IV}, does also have a different nature as compared to the XXZ spin chain, as one can see by assessing the qualitative differences between correlation functions. It turns out that the analysis of the long distance $\langle \sigma_1^z \sigma_{\ell+1}^z \rangle$ correlation suffices to our purposes. Similarly to the XXZ case \cite{KLUMPER2001,CORRXXZ1,CORRXXZ2,CORRXXZ3}, the Trotter-Suzuki decomposition can be exploited in the context of correlation function, such that,
\begin{equation}
\langle \sigma^z_1 \sigma^z_{\ell+1} \rangle_{N,L} = \frac{\tr D^{L-\ell} \cdot K^{-1} \sigma^z_j K \cdot D^{\ell} \cdot K^{-1} \sigma^z_j K}{\tr D^L} - {\left(\frac{\tr D^{L} \cdot K^{-1} \sigma^z_j K}{\tr D^L}\right)}^2,
\end{equation}
where $D$ is the diagonal matrix containing the quantum transfer matrix eigenvalues and $K$ the matrix whose columns are the corresponding eigenvectors, i.e $T^{QTM}(0)=K D K^{-1}$. If we let $1\ll \ell \ll L \rightarrow \infty$, we get
\begin{equation}
\langle \sigma^z_1 \sigma^z_{\ell+1} \rangle = \lim_{N \rightarrow \infty} \sum_{r\in \text{sub-dominant}} {\left(\frac{\Lambda_{r}(0)}{\Lambda_{\text{max}}(0)}\right)}^{\ell} \left[K^{-1} \sigma^z_j K\right]_{\text{max},r} \left[K^{-1} \sigma^z_j K\right]_{r,\text{max}}.
\end{equation}
For the long distance behavior we may restrict the above summation to the sub-dominant eigenvectors such that the form-factor is different from zero for $N \rightarrow \infty$. Whereas for the periodic XXZ spin chain this comes from the sector of zero magnetization with no degeneracy for large temperatures \cite{KLUMPER2001}, we find that the required sub-dominant eigenvalues for the IRF quantum transfer matrix come from the twisted part and has degeneracy $g=2$ for all temperatures, i.e. they come in pair of complex conjugates. Let $\Lambda_1$ be one such eigenvalue, then
\begin{equation}
\ln \frac{\Lambda_1(0)}{\Lambda_{\text{max}}(0)}= - \xi^{-1}+ \im \kappa,
\end{equation}
where $\xi$ is the correlation length and $\kappa$ the wave-vector(momentum) of the oscillation.

\begin{figure}[h]
\begin{center}
  \includegraphics[width=0.43 \linewidth, angle=-90]{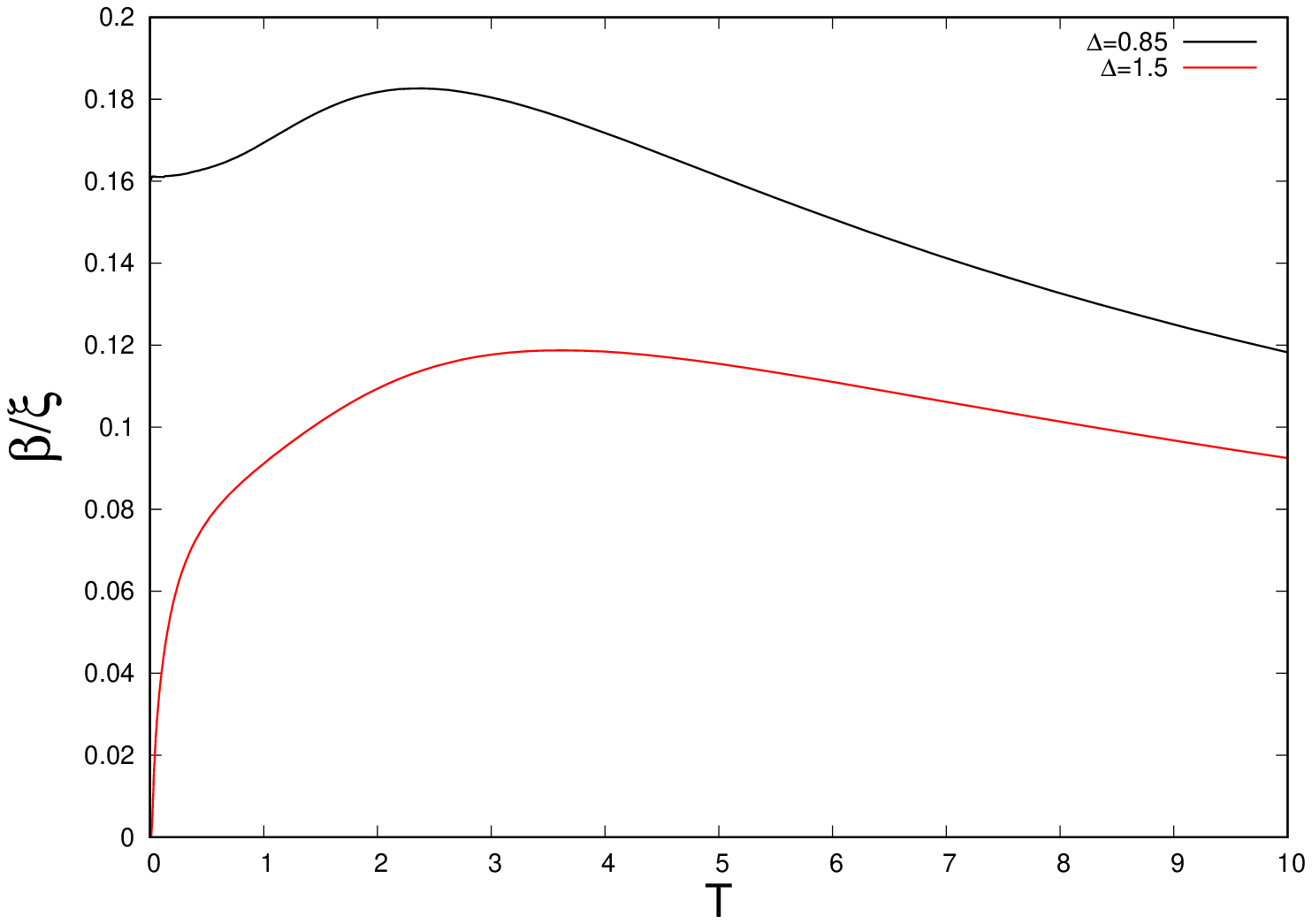}
  \includegraphics[width=0.43 \linewidth, angle=-90]{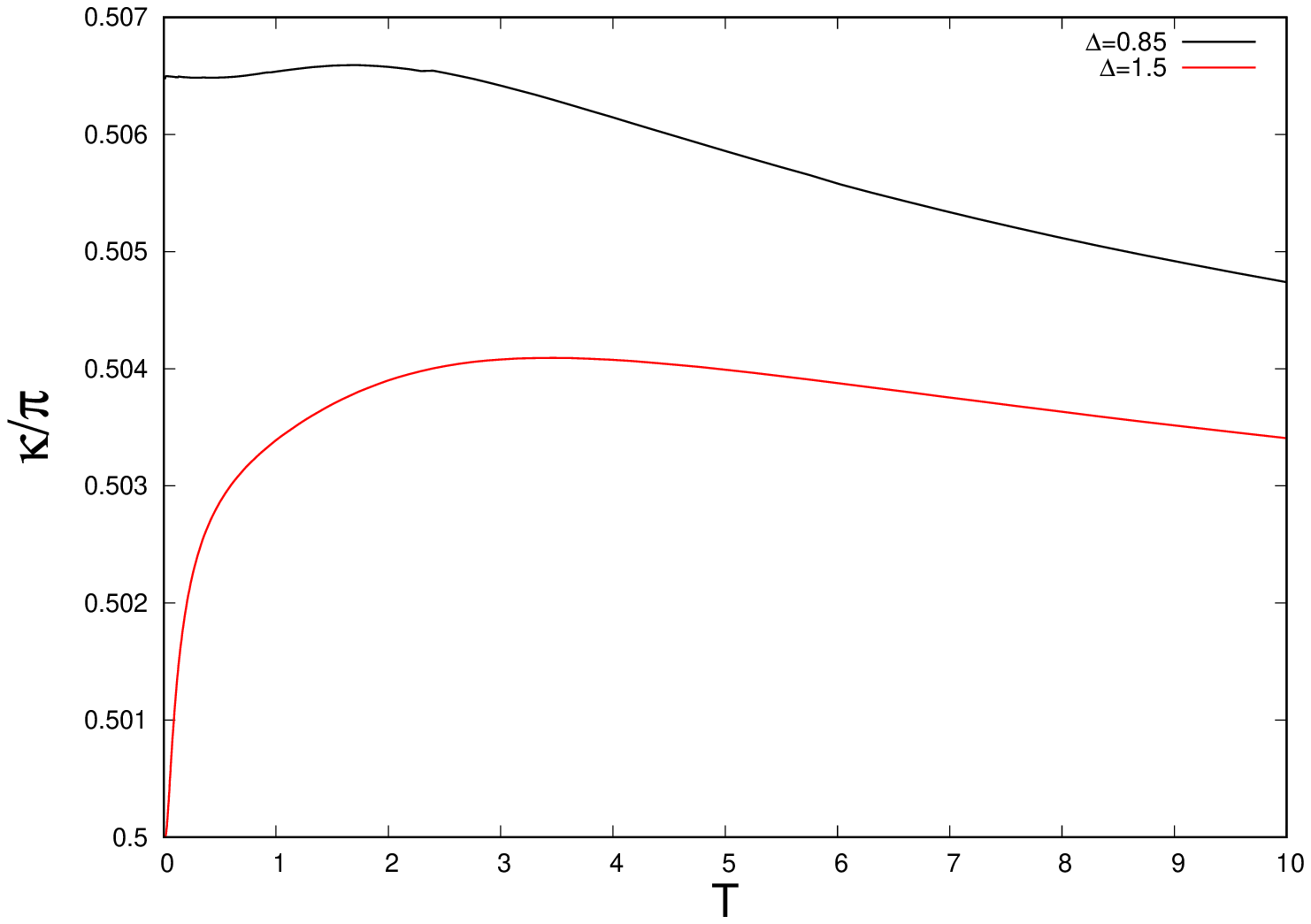}
  \caption{ a) Inverse of the correlation length $\beta/\xi$ and b) momentum of the oscillation $\kappa/\pi$ for the long distance correlation $\langle \sigma_1^z \sigma_{\ell+1}^z \rangle$ at $J=0.1$ for  $\Delta=0.85$ and $\Delta=1.5$. }
  \label{Result1}
  \end{center}
\end{figure}

\begin{figure}[htb]
\begin{center}
		\includegraphics[width=0.43\linewidth, angle=-90]{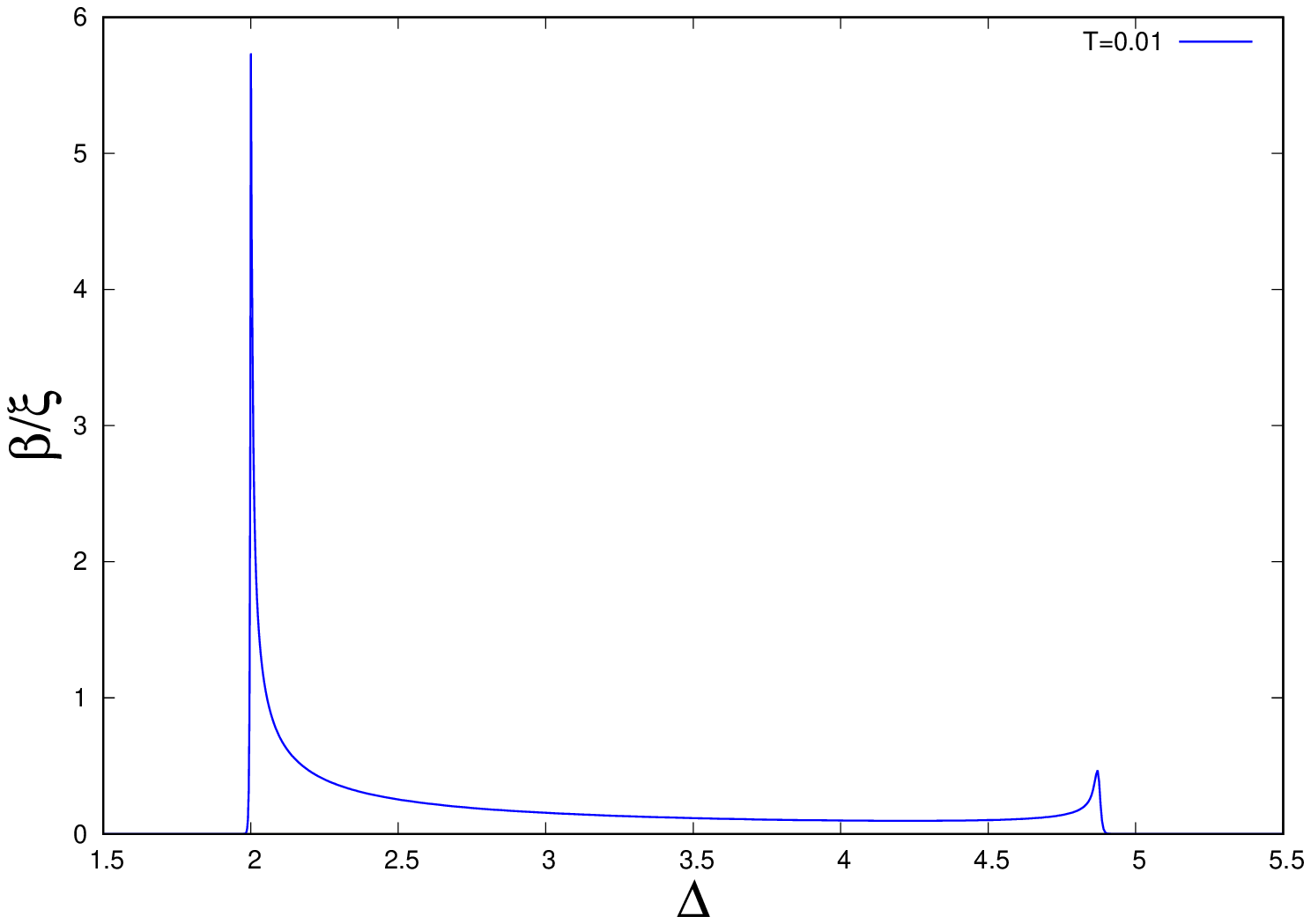}
		\includegraphics[width=0.43\linewidth, angle=-90]{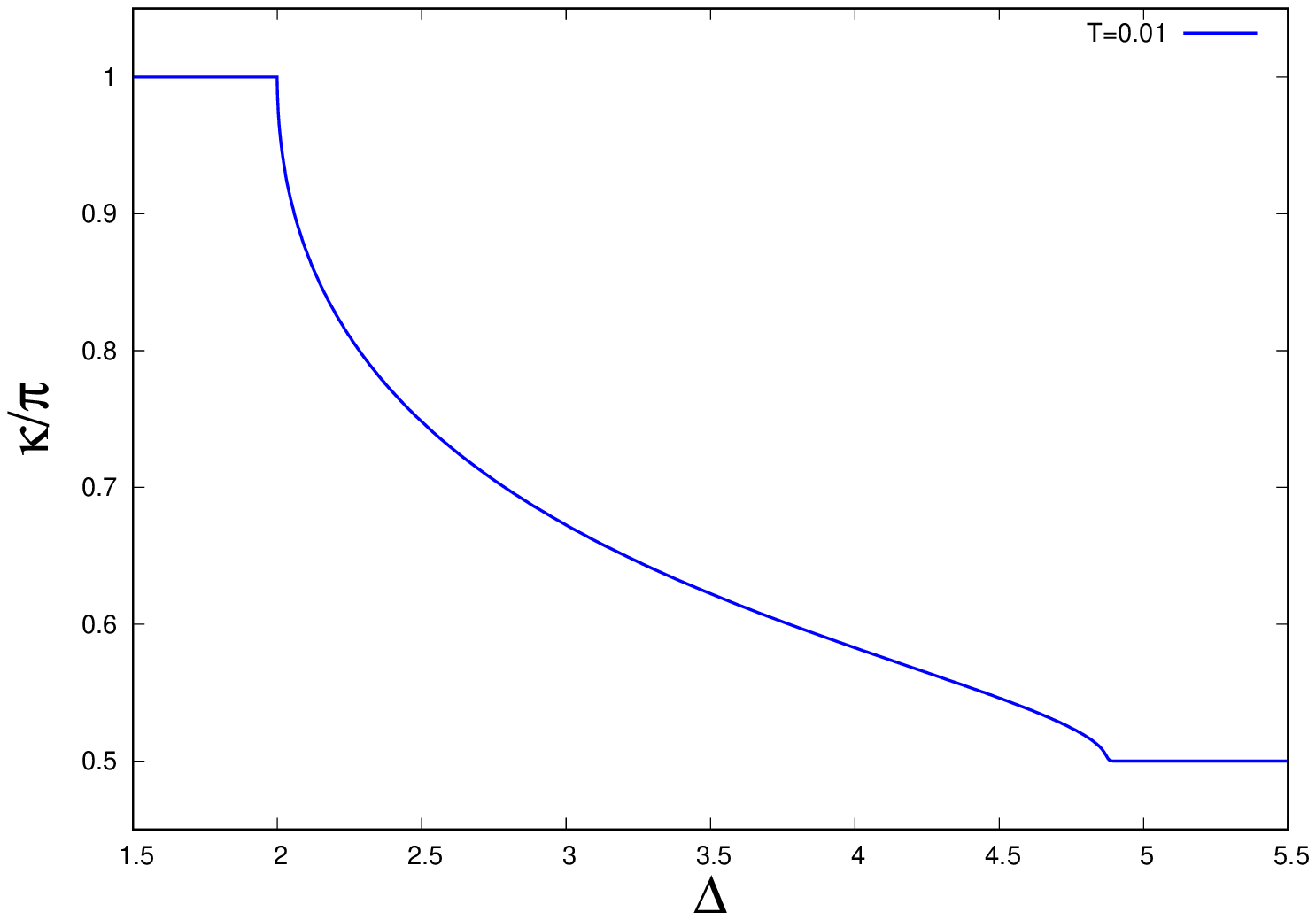}
\caption{a) Inverse of the correlation length $\beta/\xi$; b) momentum of the oscillation $\kappa/\pi$  for the long distance correlation for $J=6$ and at temperature $T=0.01$ as a function of $\Delta$. }
\label{versusDelta}
\end{center}
\end{figure}

Numerical solution of the non-linear integral equations allows us to obtain the typical behavior of this long-distance correlation, see Figures \ref{Result1} and \ref{versusDelta}.

At any finite temperature, the correlation length and the wave-vector are finite, which expresses the fact that criticality occurs precisely at zero temperature for $|\Delta|<1$.
This is typical of unitary quantum one-dimensional systems with short-ranged interactions.  Moreover, at zero temperature the correlation length diverges linearly with $\beta$, which is reminiscent of conformal symmetry for $|\Delta|<1$.

The low-temperature analytical behavior extracted from the non-linear integral equations in (\ref{k0correction}) shows that for $J=0$,
\bear
	\kappa &=& \frac{\pi}{2}, \\
	\lim_{\beta \rightarrow \infty} \frac{\beta}{\xi} &=& \frac{1}{8 (1-\frac{\gamma}{\pi})} \frac{\gamma}{\sin \gamma}, \label{xiT0}
\ear
for $|\Delta|<1$.  The $J$ dependence can also be included into the oscillatory  phase $\kappa$, which results in $\kappa=\frac{\pi}{2} +\frac{J}{2 (\pi-\gamma)} \frac{\gamma}{\sin \gamma}$. This is in good agreement with the numerical results in Figure \ref{Result1}. Unfortunately, we could not provide the $J$ dependence of the correlation length, since this would demand the computation of $1/\beta$ corrections whose analytic evaluation has eluded us so far.

We have also computed the correlation length and the momentum oscillation for $J=6$ as a function of the parameter $\Delta$, see Figure \ref{versusDelta}. This illustrates the phase transitions among the phases I, III and IV given in Figure \ref{phase-diagram}.

In addition, the mapping (\ref{MAP}) allows us to obtain the conformal central charge, which is $c=1$, and the exponents in terms of the Gaussian model\cite{KLUMPER93}. These results from the finite-size corrections of the row-to-row transfer matrix eigenvalue, which can be written as,
\begin{equation}
\ln \Lambda^{\text{row}}(x)- L g_+(x) \approx \im \pi \nu_+-\frac{2 \pi}{L}({\rm e}^{\frac{\pi x}{\gamma}} (h_+-\frac{1}{24})+{\rm e}^{-\frac{\pi x}{\gamma}} (h_--\frac{1}{24})), \label{ScaledEigenvalue}
\end{equation}
where we denote the bulk non-universal part of the eigenvalue as $g_+(x)$, $\nu_+$ is an integer number associated to the number of positive scaling zeros in the analytical strip and the exponents are given in terms of
\begin{equation}
h_{\pm} = \frac{1}{4(1-\frac{\gamma}{\pi})} {\left(M (1-\frac{\gamma}{\pi})\pm \left(E + \frac{\phi}{\pi}\right)\right)}^2. \label{6VExponents}
\end{equation}
For even system size $L$, the magnetic sector $M$  should be multiple of 2, while index $E$ can be any integer. These results are limited to the case $J=0$.

The correlation length (\ref{xiT0}) at $J=0$ agree with (\ref{6VExponents}) where the first excitation is given by $E=M=0$ and $\phi=\frac{\pi}{2}$. Besides the sound velocity is $v=\frac{2 \pi \sin \gamma}{\gamma}$. In equivalent circumstances, the periodic $XXZ$ spin chain obtains $E=1$ and $M=\phi=0$ for the exponent dictating this long distance correlation. Hence, no floating oscillatory behavior is found from the most important contribution. In order to compare, for the $XXZ$ spin chain we have that $\kappa=\pi$ and $\lim_{\beta \rightarrow \infty} \tfrac{\beta}{\xi} = \frac{1}{2 (1-\frac{\gamma}{\pi})} \frac{\gamma}{\sin \gamma}$ \cite{KLUMPER2001}. All that implies that although the spectrum of (\ref{Hirf}) can be described in terms of the twisted XXZ, all phases present different order compared to the latter, including the critical region, for which the critical exponents are a particular combination of sectors of the XXZ chain.

\section{Conclusion}\label{CONCLUSION}

We studied an integrable quantum spin chain obtained from IRF version of the six/eight-vertex model. In order to study its physical properties, we applied the quantum transfer matrix approach in the IRF language. This allowed us to determine the ground state phase diagram and also to analyze the predominant order in each phase.

On the side of the classical IRF model studied here, the model have the advantage that the parameter $\Delta$ changes continuously, which is in contrast with the fixed values in the CSOS and RSOS models. On the side of the quantum spin chain, although the associated spin chain shares part of the integrability with the $XXZ$ spin chain, they have different symmetries, which resulted in different spin ordering in the phase diagram.  While the periodic XXZ has parity, time-reversal, spin reflections $\Pi^{x,y,z}$ and cyclic translation as discrete symmetries, when directly analyzed, our spin model does not possess time-reversal, due to the single spin ($\sigma^x$ Zeeman term) and the three spins interactions. Moreover, only the spin reflection $\Pi^x$ is a symmetry. However, when including each of these two models with their corresponding $U(1)$ charges, while this explicitly breaks time-reversal and leaves only spin reflection $\Pi^z$ for the XXZ, no explicit break occurs in our model, as time-reversal is manifestly broken and the spin reflection $\Pi^x$ continues to be a symmetry. This is because the $U(1)$ charge, in this case, is an interaction of the Ising type.

The shape of the phase diagram is the same as the XXZ spin chain, as the largest quantum transfer matrix eigenvalue are the same in either case. However, our analysis shows that not only a numerical difference on the long-distance correlation function $\langle \sigma^z_1 \sigma^z_{\ell+1} \rangle$ exists, inside the critical region. Qualitatively, the long-distance correlation at $J=0$ and at zero temperature oscillates at a wave vector locked at $\frac{\pi}{2}$, while the same does not happen for the Heisenberg spin chain\cite{KLUMPER2001}. For $J\neq 0$, the wave-vector start to vary continuously.  On the other hand, the gapped phases of (\ref{Hirf}) also differ from the XXZ chain, as it dimerizes in presence of frustration.

We expect that new quantum spin chain can be obtained from other IRF models related to more general vertex models. We also expect that correlation functions can be evaluated via functional equations of the quantum Knizhnik-Zamolodchikov type \cite{FRAHM1,FRAHM2}.  We hope to address theses problems in the future.

\section*{Acknowledgments}

TST thanks for support of the Institut Universitaire de France and the European Research Council (advanced grant NuQFT No 669205) and hospitality of the CEA/Saclay where this work started. TST would like to thank M.J. Martins for bringing the IRF models to his attention and for discussions. GAPR thanks F.C. Alcaraz for discussions and FAPESP (grant number 2023/03947-0) for funding.

\end{document}